\documentclass[preprint,12pt]{elsarticle}

\usepackage{amssymb}
\usepackage[utf8]{inputenc}
\usepackage{soul}
\usepackage{color}
\usepackage{graphicx}
\usepackage{amsmath}
\usepackage[version=4]{mhchem}
\usepackage{siunitx}
\usepackage{longtable,tabularx}
\usepackage{caption}
\usepackage{float}
\usepackage[nameinlink,capitalise,noabbrev]{cleveref}
\usepackage[dvipsnames]{xcolor}
\usepackage{algorithm}
\usepackage{algorithmic}
\graphicspath{{Figures/}}
\usepackage{subfigure}
\usepackage{multirow}
\usepackage{booktabs}
\usepackage{xcolor}

\begin{document}

\begin{frontmatter}

\title{Predicting Transonic Flowfields in Non--Homogeneous Unstructured Grids Using Autoencoder Graph Convolutional Networks}

\author[inst1,inst2]{Gabriele Immordino\corref{cor1}\fnref{label2}}
\ead{G.Immordino@soton.ac.uk}

\fntext
[label1]{Ph.D. Student}
\cortext[cor1]{Corresponding Author}

\affiliation[inst1]{organization={Faculty of Engineering and Physical Sciences, University of Southampton},
            city={Southampton},
            country={United Kingdom}}

\author[inst2]{Andrea Vaiuso\fnref{label2}}
\fntext[label2]{Research Associate}

\author[inst1]{Andrea Da Ronch\fnref{label3}}
\fntext[label3]{Associate Professor, AIAA Senior Member}

\author[inst2]{ Marcello Righi\fnref{label4}}
\fntext[label4]{Professor, AIAA Member, Lecturer at Federal Institute of Technology Zurich ETHZ}

\affiliation[inst2
]{organization={School of Engineering, Zurich University of Applied Sciences ZHAW},
            city={Winterthur},
            country={Switzerland}}

\begin{abstract}

This paper focuses on addressing challenges posed by non--homogeneous unstructured grids, commonly used in Computational Fluid Dynamics (CFD). Their prevalence in CFD scenarios has motivated the exploration of innovative approaches for generating reduced--order models. The core of our approach centers on geometric deep learning, specifically the utilization of graph convolutional network (GCN). The novel Autoencoder GCN architecture enhances prediction accuracy by propagating information to distant nodes and emphasizing influential points. This architecture, with GCN layers and encoding/decoding modules, reduces dimensionality based on pressure--gradient values. The autoencoder structure improves the network capability to identify key features, contributing to a more robust and accurate predictive model. To validate the proposed methodology, we analyzed two different test cases: wing--only model and wing--body configuration. Precise reconstruction of steady--state distributed quantities  within a two--dimensional parametric space underscores the reliability and versatility of the implemented approach.

\end{abstract}

\end{frontmatter}

\section*{Nomenclature}
\textbf{Acronyms} 
{\renewcommand\arraystretch{1.0}
\noindent\begin{longtable*}{@{}l @{\quad=\quad} l@{}}

$AE-GCN$ & autoencoder graph convolutional network \\
$CFD$ & computational fluid dynamics \\
$GCN$ & graph convolutional network \\
$GNN$ & graph neural network \\
$LHS$ & Latin Hypercube Sampling \\
$ML$ & machine--learning \\
$MAE$ & mean absolute error \\
$MAPE$ & mean absolute percentage error \\
$MSE$ & mean squared error \\
$MWLSI$ & moving weighted least squares interpolation \\
$ROM$ & reduced--order model \\

\end{longtable*}}

\textbf{Symbols} 
{\renewcommand\arraystretch{1.0}
\noindent\begin{longtable*}{@{}l @{\quad=\quad} l@{}}

$AoA$   & angle of attack, deg \\
$c$   & mean chord, m \\
$C_{D}$ & drag coefficient \\
$C_{F}$ & skin friction coefficient \\
$C_{L}$ & lift coefficient \\
$C_{My}$ & pitching moment coefficient \\
$C_{P}$ & pressure coefficient \\
$M$ & Mach number \\
$Re$ & Reynolds number \\
\end{longtable*}}

\section{Introduction}

In recent years, addressing problems characterized by non--homogeneous and unstructured grids has become a central topic of research in the field of aerospace engineering. A pertinent example lies within the Computational Fluid Dynamics (CFD) field, where the initial step involves the mesh generation, entailing the discretization of the fluid domain through the finite volume method. This mesh serves as a computational grid that enables the simulation of fluid flow and related phenomena within a defined space. A non--homogeneous unstructured grid is characterized by irregularly shaped elements (such as triangles or tetrahedras) connected in a non-regular pattern. The spacing between grid points varies across the domain, providing greater resolution in areas of interest, such as regions with complex geometries or flow features, while optimizing computational resources in less critical areas.

The complexities inherent in non--homogeneous unstructured geometries, especially when predicting intricate fluid flow scenarios have given rise to a pressing need for innovative approaches in generating reduced--order models (ROMs). Within this context, machine learning has emerged as a promising avenue to tackle the challenges posed by these non-traditional data structures. Initial efforts in this domain centered around the application of deep neural networks, demonstrating their efficacy in capturing intricate patterns and relationships within the fluid dynamics domain~\cite{sabater2022fast,castellanosassessment,immordino2023steady}. Nevertheless, as the complexity of non--homogeneous unstructured grids became more apparent, the necessity for a more sophisticated architecture became evident. 


The concept of geometric deep learning emerged around 2017~\cite{bronstein2017geometric}, introducing the use of graph-structured data prediction through the adoption of graph neural network (GNN) architectures~\cite{gori2005new}. Specifically designed for applications involving interconnected entities, GNNs excel in capturing intricate relationships and dependencies within graph nodes and connections between nodes~\cite{wu2020comprehensive,zhang2020deep,zhou2020graph}. The inherent ability of GNNs to consider both local and global context through neighborhood aggregation mechanisms makes them well-suited for tasks where topological information is critical. These versatile networks have found extensive application as a foundation for solving classical artificial intelligence tasks and addressing various challenges in data science and analysis~\cite{wu2020comprehensive}. This includes applications such as social influence prediction~\cite{qiu2018deepinf}, prevention of adversarial attacks~\cite{zugner2018adversarial}, electrical health records modeling~\cite{choi2017gram}, analysis of brain networks~\cite{kawahara2017brainnetcnn}, and event detection~\cite{nguyen2018graph}. Notably, it has been shown that GNNs outperform traditional approaches in handling local nonlinearities~\cite{hines2023graph}. They have demonstrated precise predictions for aerodynamic performances~\cite{juangphanich2023predicting} and flowfield properties~\cite{ogoke2021graph}. Additionally, they result effective in addressing complex time-dependent problems~\cite{massegur2023recurrent} and have proven successful in diverse aerospace applications, including data fusion tasks~\cite{li2023multi}, uncertainty quantification~\cite{li2023uncertainty} and multi-objective optimization~\cite{li2024accelerating}.

\begin{figure}[b]
    \centering
    \includegraphics[width=1\linewidth]{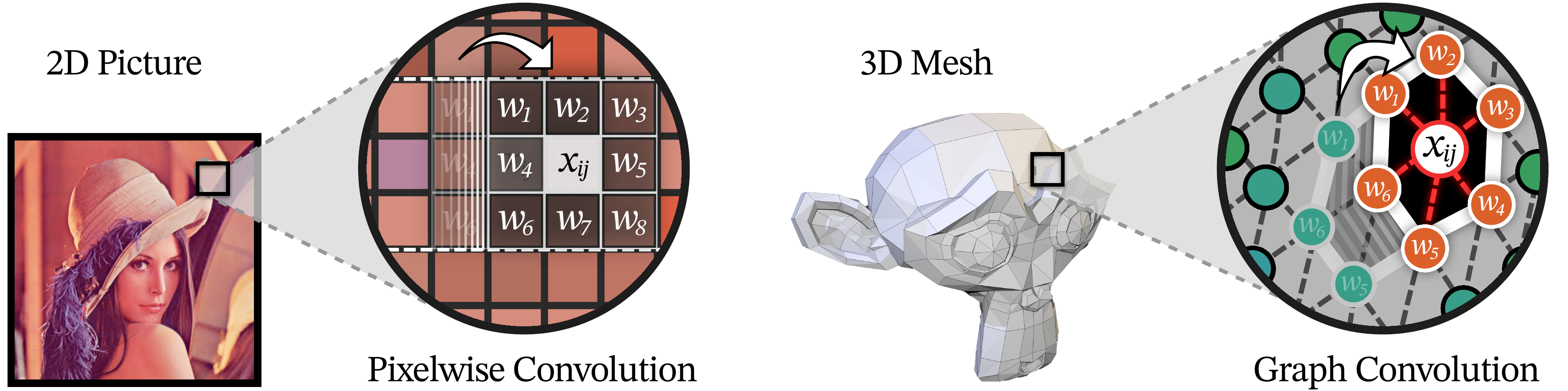}
    \caption{Visual comparison between pixelwise convolution on a 2D digital image and graph convolution on a 3D mesh.}
    \label{fig:conv}
\end{figure}

While Convolutional Neural Networks (CNNs) have demonstrated remarkable accuracy across various domains~\cite{jin2018prediction,fukami2019super,omata2019novel,peng2020time,rozov2021data}, they rely on the assumption that inputs exhibit a Cartesian grid structure. This assumption allows CNNs to leverage three fundamental properties—sparse connection, parameter sharing, and translation invariance—to achieve accurate results. However, this limitation confines CNNs to regular grid data, such as images (2D grids) and texts (1D sequences). Consequently, our approach involves the adoption of Graph Convolutional Networks~\cite{duvenaud2015convolutional}, which harness the convolutional operation of CNNs and extend it to non--homogeneous unstructured data. It involves a single-element filter swept across the connected nodes and being weighted by the corresponding edge weights, hence the convolutional analogy (refer to Figure~\ref{fig:conv}). This idea enables the application of convolutional operations to data structures without the regular grid assumption, broadening our predictive capabilities and allowing direct input of raw 3D model mesh data to GCNs. This approach avoids unnecessary pre--computation or feature extraction methods that may introduce bias or loss of information. 

We adopt the methodology introduced by Massegur et al.~\cite{massegur2023graph} to propagate information to nodes located farther away which proved effective in analogous scenarios. This methodology involves the autoencoder GCN architecture, which enhances prediction accuracy by establishing intricate connections within the reduced--space, comprising only the most influential points in the solution. 

Our architecture is specifically designed with GCN layers, complemented by pooling and unpooling layers, that effectively reduces and expands the dimensionality of the latent spaces in accordance with the pressure--gradient values and propagates information to nodes located further away. This integration enhances the network ability to discern and emphasize key features, ultimately contributing to a more robust and accurate predictive model. Two test cases are proposed to validate the developed methodology: wing--only model and wing--body configuration.

This study introduces the adoption of a dimensionality reduction module based on the pressure--gradient values, the implementation of a fast connectivity reconstruction employing the Mahalanobis distance, Bayesian optimization of network architecture, exploration across two test cases characterized by distinct physical phenomena, and the integration of a physics--informed loss function incorporating a penalty term for pitching moment coefficient. Together, these contributions yield a systematically lower calculated error compared to prior studies.

The structure of the paper is as follows: Section~\ref{sec:methodology} outlines the methodology implemented, where a comprehensive explanation of the architecture and its blocks is given, Section~\ref{sec:results} presents the results obtained on examples of steady--state prediction of aircraft wing configurations, and Section~\ref{sec:Conclusions} summarises the conclusions drawn from the study.

\section{Methodology}\label{sec:methodology}

This section explains the methodology that guided the creation of the model at hand. Initially, the general autoencoder graph convolutional network architecture is introduced, followed by a detailed explanation of each component that constitutes every module of the model.

\subsection{Graph Autoencoder Architecture}

\begin{figure}[!htb] 
    \centering
    \includegraphics[width=\textwidth]{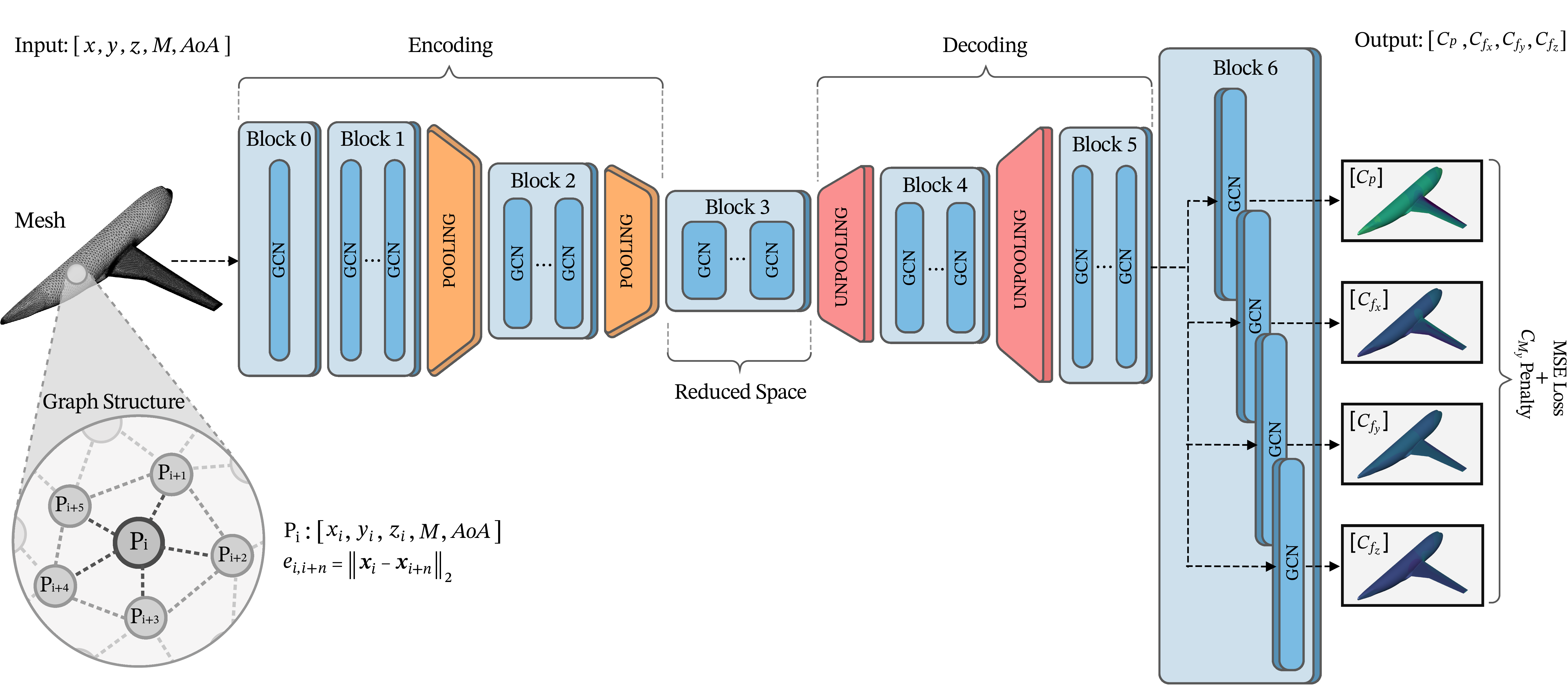}
    \caption{Schematic of the graph autoencoder architecture.}
    \label{fig:architecture}
\end{figure}

The steady--state prediction ROM developed in this work uses freestream conditions and mesh coordinates as input and is designed to predict specific values for each point in the graph. Scalar freestream conditions are assigned to each node of the surface alongside their respective coordinates. An Autoencoder GCN model (AE--GCN) with two levels of dimensional reduction/expansion, involving custom pooling/unpooling layers, was implemented. The output of the model is generated by four parallel GCN layers. The whole architecture is finally trained for the pointwise prediction of the four desired output $C_p,C_{f_{x}},C_{f_{y}}$ and $C_{f_{z}}$. A schematic of the model architecture is illustrated in Figure~\ref{fig:architecture}.

 

The use of an Encoder-Decoder based architecture aims to reduce the computational effort by reducing the size of the data during the prediction, increasing the scalability of the system, and also allows the model to consider the connection between more distant points of the mesh, which are not directly connected initially. This step has been taken in order to reproduce the CNN behaviour used in AI-based computer vision tasks~\cite{gu2018recent,li2021survey}, with the addition of the information about the distances and connections between the points given by the graph structure.

The pooling module implemented in our approach is a gradient--based point selection and connection reconstruction. The pressure gradient--based point selection task involves two key steps. Firstly, we compute the gradients for each sample and subsequently identify the regions of interest across all samples. This process enables us to pinpoint areas where pressure gradients exhibit significant disparities, thereby identifying points characterized by heightened nonlinearity. Once these critical points are identified, we implement a Moving Weighted Least Squares Interpolation (MWLSI) algorithm~\cite{quaranta2005conservative,joldes2015modified} to seamlessly interpolate values from the source points (fine grid) to the destination points (corase grid). To reconstruct connections and calculate Euclidean distances between the remaining points, a Mahalanobis distance-based method was implemented~\cite{de2000mahalanobis}. This method re-establishes connections of each point with its 5 neighbors in the destination space based on Mahalanobis distances calculated in the original space.

The aim of the unpooling module is to reconstruct the original structure of the input to generate an output with the same dimension of the input, but this operation requires a new interpolation matrix computed with the MWLSI algorithm (refer to Section~\ref{subsec:MWLS} for details) to calculate the missing data of the new nodes, moving from a coarser grid back to the finer one. The pooling and unpooling modules are pre-computed in order to save computational resources. An on demand version could be implemented for adding learnable capabilities of space reduction/expansion, especially on time--variant problems.

To enhance the predictive capacity of the model, we adopted two strategies: a Bayesian optimization and a custom loss function. The Bayesian approach has been employed for optimizing the neural network hyperparameters, such as number of layers per block, units per layers and compression ratio of encoding/decoding operations. By leveraging Bayesian optimization, the model systematically explores and adapts these hyperparameters to maximize performance and predictive accuracy. The custom loss function aims to optimize the distribution of $C_P$ and $C_F$ components across the grid by minimizing the mean squared error (\texttt{MSE}) between the model predictions and the ground truth. Factors like shock waves and boundary layer separation introduce complexity to predictions, affecting force resultant and, therefore, moment calculation. To address this, a penalty term for the pitching moment coefficient $C_{M_y}$ has been introduced into the \texttt{MSE} loss function. This addition, represented as $Loss = \text{MSE} + \lambda \cdot C_{M_y}$, with $\lambda = 0.01$ for dimensional consistency, guides the model towards more precise predictions, particularly in terms of shock wave positioning.

\subsection{Graph Deep--learning Model}

Graph Neural Networks (GNNs) are a class of neural networks designed to work with graph-structured data. Graphs consist of nodes and edges, where nodes represent entities and edges represent relationships or connections between these entities. GNNs have gained popularity for their effectiveness in tasks involving graph-structured data.

Initially, this section introduces the representation of the wing surface mesh as a graph. Subsequently, attention is directed towards the Graph Convolutional Network (GCN).

\subsubsection*{Graph Representation}

\begin{figure}[!htb]
\centering

\subfigure[Cyclic]{\label{subfig:cyclic}
\includegraphics[trim=0 0 0 0, clip, width=0.45\linewidth]{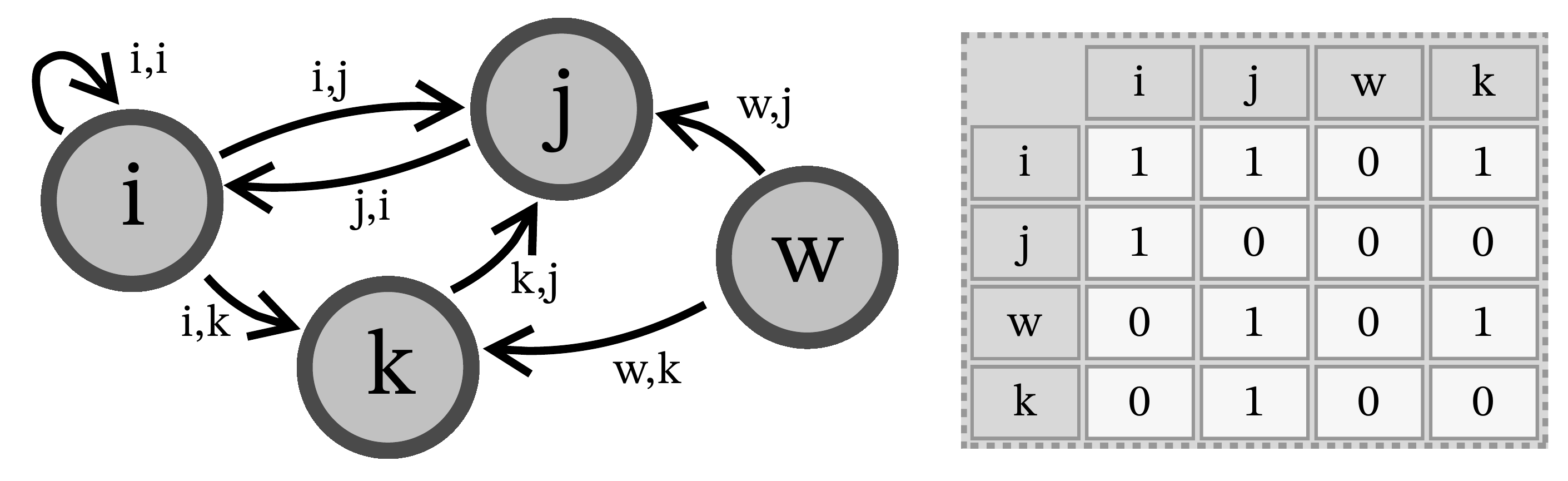}} 
\hfill
\subfigure[Acyclic]{\label{subfig:acyclic}
\includegraphics[trim=0 0 0 0, clip, width=0.45\linewidth]{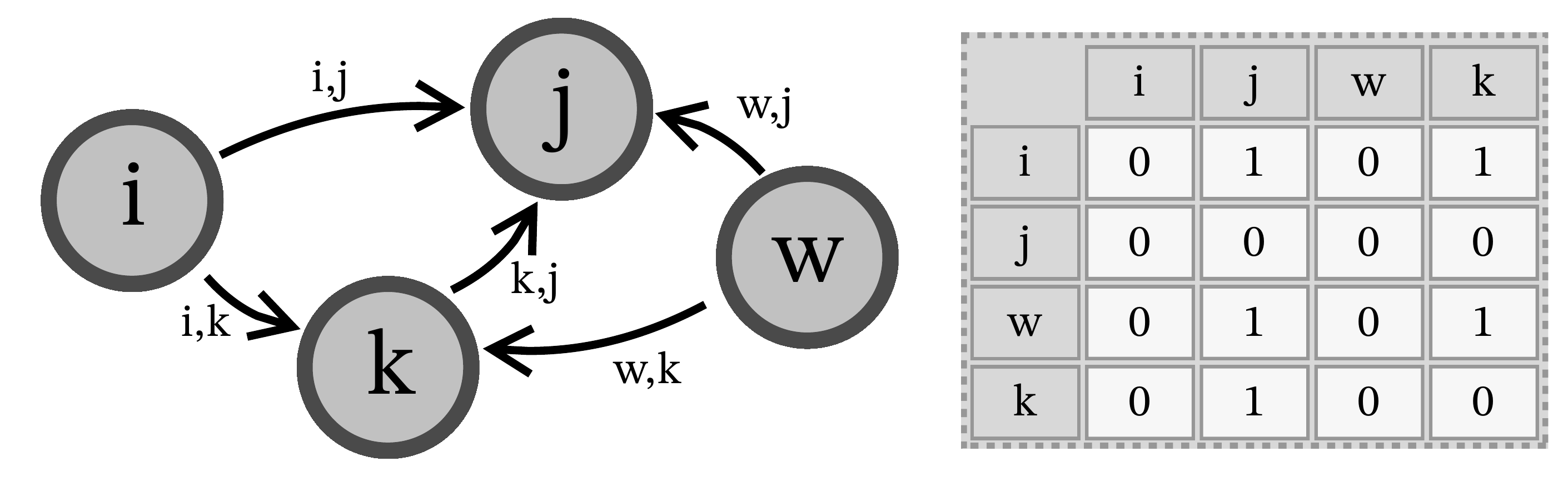}} \\

\caption{Visual representation of a graph diagram and its connectivity matrix.}
\label{fig:graph_comparison}
\end{figure}

A graph $G$ consists of nodes ($N$) and edges ($E$). Edge $(i,j)$ denotes directional connection from node $i$ to node $j$, differing from $(j,i)$ when $i\neq j$. Self-loops are possible if $(i,i)\in E$. Graphs are commonly illustrated graphically using circles for nodes and arrows for connections. In the graph $G$ in Figure~\ref{fig:graph_comparison} with nodes $N=\{i,j,k,w\}$, edges between nodes are represented by one-way arrows. Connections can be expressed through a matrix notation, where $\mathbf{A}_{ij} = 1$ if $(i,j)\in E$, and $\mathbf{A}_{ij} = 0$ otherwise. This matrix, called \textit{adjacency matrix} or \textit{connectivity matrix}, may become significantly sparse with a large number of nodes. Numerous techniques exist for storing adjacency matrices efficiently. Graphs can also carry edge costs, denoted as $e_{ij}$, representing distances or other values, including negatives. In an adjacency matrix for a graph with costs, replace $1$ with the cost and use $\infty$ for absent connections. A path $p(i\rightarrow j)$ in a graph is a finite series of steps $\langle n_k,n_{k+1}\rangle$ for $k=0,1,...,K$, where $n_k \in N$, $(n_k,n_{k+1}) \in E$, $n_0 = i$ and $n_K = j$. A graph $G$ is defined \textit{acyclic} only if $\forall i \in N$ there are no paths $p(i\rightarrow j)$ where $i=j$, otherwise is \textit{cyclic}. Examples of cyclic and acyclic graphs are depicted in Figure~\ref{subfig:cyclic} and \ref{subfig:acyclic}, respectively.

In our context, the mesh could be considered as a cyclic graph \(G\) wherein each grid point \(i\) in the surface mesh \(G\) is a node characterized by variables (features), which are positional coordinates \(\mathbf{x}_i\), pressure coefficient \(C_{P_i}\) and three components of skin fiction coefficient \(C_{F_i}\). The connections between grid points form the edges of the graph, linking target node \(i\) with grid points \(j \in S\). The nodes features are denoted as \(y_i\), and the weights on edges are denoted as \(e_{ij}\).

Graph connectivity is expressed through the adjacency matrix \(\mathbf{A}\), where each entry \(e_{ij}\) represents the weight on the edge connecting node \(j\) to node \(i\). The weights are determined by the Euclidean distance between adjacent grid points: \(e_{ij} = \|\mathbf{x}_i - \mathbf{x}_j\|_2\). To normalize the edge weights within the range \((0, 1]\), including self-loops with \(e_{ii} = 1\), the adjacency matrix is augmented by the identity matrix: \(\hat{\mathbf{A}} = \mathbf{A} + \mathbf{I}\). Additionally, since $\forall (i,j) \in E \, \exists (j,i)\in E $ and \(e_{ij} = e_{ji}\), the adjacency matrix results symmetric: \(\hat{\mathbf{A}} = \hat{\mathbf{A}}^T\).

Considering the sparsity of both the graph connectivity and the adjacency matrix, a more memory-efficient organization in Coordinate List (COO) format is adopted. The edge-index matrix has dimensions \(n_e \times 2\) (pairs of node indices), and the edge-weight matrix is \(n_e \times 1\), where \(n_e\) represents the number of edges in the mesh.

\subsubsection*{Graph Convolutional Network}

In this study, we chose to employ GCN layers based on the graph convolutional operator. This operator was introduced by Duvenaud et al. in 2015~\cite{duvenaud2015convolutional} for extracting features from molecular fingerprints. Kipf et al. extended this work in 2016~\cite{kipf2016semi}, providing the foundation for the current implementation in the \texttt{PyTorch-Geometric} Library~\cite{Fey/Lenssen/2019} used in this paper. GCNs are renowned for their ability to generate node embeddings that capture essential structural information on a graph. This is particularly beneficial for tasks that necessitate an understanding of relationships and connections between entities. GCNs utilize a convolutional operation similar to classical CNNs to aggregate information from neighboring nodes, while also incorporating distance information from the local neighborhood. The scalability of GCNs is facilitated by parameter sharing, as the parameters are uniformly shared across all nodes.


The GCN operator follows the layer-wise propagation rule that is defined by:

\begin{equation}\label{eq:gcn}
    H^{(l+1)}=\sigma (\mathbf{\Tilde{D}}^{-\frac{1}{2}}\mathbf{\Tilde{A}}\mathbf{\Tilde{D}}^{-\frac{1}{2}}H^{(l)}W^{(l)})
\end{equation} 

Where $H^{(l)}$ denotes the input graph at layer $l$ and $H^{(l+1)}$ represent the output at layer $l+1$. The matrix $\mathbf{\Tilde{A}}=\mathbf{A}+\mathbf{I}_N$ represents the adjacency matrix with added self-loops to each node. The matrix $\mathbf{\Tilde{D}}$ is a diagonal matrix defined as $\mathbf{\Tilde{D}}_{ii}=\Sigma_j(\mathbf{\Tilde{A}}_{ij})$. The trainable matrix specific to the layer is denoted as $W^{(l)}$ and $\sigma$ denotes the application of an activation function. Equation~\eqref{eq:gcn} is motivated via a first-order approximation of trainable localized spectral filters $g_{\theta}$ on graphs~\cite{kipf2016semi}.


A spectral convolution (denoted by $\ast$) of an input graph $\Tilde{x}$ with a filter $g_{\theta}$ parametrized by $\theta$ in the Fourier domain can be defined as:

\begin{equation}
    g_{\theta} \ast x = \mathbf{U} g_{\theta} \mathbf{U}^T x
\end{equation}

Here, $U$ represents the matrix of eigenvectors, with its eigenvalues denoted as $\Lambda$, obtained from $\mathbf{L} = \mathbf{I}_N - \mathbf{D}^{-\frac{1}{2}}\mathbf{A}\mathbf{D}^{-\frac{1}{2}} = \mathbf{U\Lambda U}$, where $\mathbf{D}_{ii} = \sum_j (\mathbf{A}_{ij})$ is a diagonal matrix. By expressing $g_{\theta}$ as a function of $\Lambda$ and approximating it through a truncation of Chebyshev polynomials up to the $K^{th}$ order~\cite{hammond2011wavelets}, the eigen-decomposition of $\mathbf{L}$ can be easily computed, resulting in:

\begin{equation}\label{eq:conv}
    g_{\theta} \ast x \approx \sum_{k=0}^{K}{\theta'_k T_k (\mathbf{\Tilde{L}}) \, x}
\end{equation}

Where $\theta'$ is a vector of Chebyshev coefficients, and $T_k (\mathbf{\Tilde{L}})$ is the $k^{th}$ Chebyshev polynom applied to $\mathbf{\Tilde{L}}=\frac{2}{\lambda_{max}}\mathbf{L}-I_N$ with $\lambda_{max}$ denoting the maximum eigenvalue of the matrix $\mathbf{\Lambda}$.

Reducing the number of parameters is beneficial for addressing overfitting and streamlining operations per layer. By constraining the Chebyshev order to $K=1$ and approximating the value of $\lambda_{max}$ to $2$ (assuming neural network parameters adjust to this scale change during training), Equation~\eqref{eq:conv} simplifies to:

\begin{equation}\label{eq:semp_conv}
    g_{\theta} \ast x \approx \theta (\mathbf{I}_N+\mathbf{D}^{-\frac{1}{2}}\mathbf{A}\mathbf{D}^{-\frac{1}{2}})x
\end{equation} 

Repeated application of this operator can lead to numerical instabilities, causing either exploding or vanishing gradients, particularly in the context of deep neural network models. To address this issue, the use of the \textit{renormalization trick} is recommended, as reported by Kipf et al.~\cite{kipf2016semi}.

Through successive application of pooling operations, information from a node is propagated through increasingly distant neighborhoods. For instance, with $k_l$ concatenated GCN layers, we extend influence to the $k_l^{th}$-order neighborhood surrounding node $i$.

Lastly, the output of the GCN layer is fed through an activation function $\sigma$ to introduce nonlinearities. Thus, the operation at each layer $l$ consists of the GCN operator in Equation~\eqref{eq:gcn} with the Rectified Linear Unit (PReLU)~\cite{he2015delving} operator used as an activation function:

\begin{equation}
    f_a(\mathbf{y}) = \begin{cases}
                        \mathbf{y} & \text{if } \mathbf{y} \geq 0 \\
                        \beta \mathbf{y} & \text{if } \mathbf{y} < 0
                    \end{cases} \label{eq:PReLU_activation}
\end{equation}

where $\beta$ is a learnable parameter that is distinct for each channel of the input vector. A neural network model based on graph convolutions can therefore be built by stacking multiple convolutional layers defined as before.

ADAptive Moment estimation (\texttt{Adam})~\cite{kingma2014adam} was adopted during the back--propagation phase for optimising neural network weights and minimising \texttt{MSE} loss function. An adaptable learning rate has been used, starting from 0.001 and applying a learning rate decay of a factor of 0.9 every 30 epochs. A batch size equals to 1 led to the most accurate results. The ROM was implemented in the deep--learning python library \texttt{PyTorch}~\cite{NEURIPS2019_9015}, leveraging the GCN layer from \texttt{PyTorch-Geometric}~\cite{Fey/Lenssen/2019}. 


\subsubsection*{Bayesian Optimization for Hyperparameters Tuning}

To improve the predictive ability of the model, it is crucial to select an appropriate set of hyperparameters. An optimisation algorithm capable of exploring the large design space is essential. To address this, a hyperparameter tuning approach based on Bayesian optimization~\cite{snoek2012practical} was performed for each test case. The methodology employed follows the approach outlined by Immordino et al.~\cite{immordino2023steady} but is extended to accommodate our more complex architecture.

The advantage of Bayesian optimization lies in its ability to iteratively refine hyperparameters guided by Bayesian probability distribution functions, rather than exhaustively exploring all possible combinations. Each iteration, called trial, entails training the network with a specific set of hyperparameters, optimizing them based on the performance of preceding trials in terms of the validation set metric. This process continues until the optimal result is achieved. The reader is referred to Immordino et al.~\cite{immordino2023steady} for a complete insight into the method. The pseudo-code of the Bayesian optimization strategy is depicted in Algorithm~\ref{alg:bayesian_optimization_kt}.

\begin{algorithm}[!b]
\caption{Bayesian Optimization for Hyperparameter Tuning}
\label{alg:bayesian_optimization_kt}
\begin{algorithmic}[1]
\STATE \textbf{Input:} Objective function $f$, search space $S$, hyperparameter tuner $T$, number of trials $N_{\text{trials}}$
\STATE \textbf{Output:} Optimal hyperparameter set $\theta^*$
\STATE Initialize hyperparameter tuner $T$ with search space $S$
\STATE Conduct an initial random search to populate $T$

\STATE Initialize optimal hyperparameter set: $\theta^* \leftarrow \text{None}$
\STATE Initialize optimal objective value: $y^* \leftarrow -\infty$

\FOR{$i = 1$ to $N_{\text{trials}}$}
    \STATE Sample the next hyperparameter set from $T$: $\theta_i \leftarrow \text{sample}(S)$ 
    \STATE Evaluate the objective function: $y_i = f(\theta_i)$
    
    \IF{$y_i > y^*$}
        \STATE Update the optimal hyperparameter set: $\theta^* \leftarrow \theta_i$
        \STATE Update the optimal objective value: $y^* \leftarrow y_i$
    \ENDIF
    
    \STATE Update the posterior distribution conditioned on observed data
    
    \STATE Set: $\theta_{\text{next}} \leftarrow \arg\max_{\theta} \text{AcquisitionFunction}(\theta; \text{Posterior})$
    
    \STATE Add $\theta_{\text{next}}$ into $T$
\ENDFOR

\STATE \textbf{Return:} Optimal hyperparameter set $\theta^*$
\end{algorithmic}
\end{algorithm}

The design parameters targeted for the optimization process include:

\begin{itemize}
    \item \textbf{Number of layers per block}: A block defines a group of layers before or after a spatial reduction operation in the encoding module. The decoding module mirrors this structure for saving computational resources.
    \item \textbf{Number of units per layer}: This denotes the number of neurons implemented in a single GCN module. The optimization process explores this parameter only for the encoding phase and then mirrors it for the decoding phase, starting from the midpoint of the reduced space block. This approach minimizes computational cost and ensures dimensional compatibility of the layers. 
    \item \textbf{Dimensionality compression/expansion value}: This represents different compression/expansion ratios between the number of points in coarser and finer meshes during compression, and vice versa during expansion.
    
\end{itemize}

The design space for hyperparameters, including the possible values and step size for each variable, is presented in Table~\ref{tab_Hyperparameters_design_space}. The chosen ranges were intentionally set to be sufficiently large. Indeed, throughout the optimization process, it was observed that the hyperparameters converged to values below the upper limit of the specified ranges.

\begin{table}[!]
\centering
\begin{tabular}{l c c }
\hline 
\hline 
\textbf{Hyperparameter} & \textbf{Value} & \textbf{Step size} \\ 
\hline
Compression ratio   & 1/4 - 1/3 - 1/2     &  --  \\  
Number of Hidden Layers per Block   & 1 to 3   &  1 \\  
Number of Neurons per Hidden Layer  & 32 to 512   & 16 \\  

\hline
\hline 
\end{tabular}
\caption{Hyperparameters design space.}
\label{tab_Hyperparameters_design_space}
\end{table}

In our study, we used \texttt{Optuna}~\cite{akiba2019optuna}, a Python library designed for working on \texttt{PyTorch} framework, which seamlessly integrates Bayesian optimization into the hyperparameter search process. To ensure sufficient convergence towards the optimal set of hyperparameters, we conducted 30 trials. Concurrently, we imposed a constraint on the optimization algorithm by limiting each trial to 500 epochs, thus managing computational demands effectively. Upon completion of the optimization phase, we executed the training procedure for the refined encoder-decoder architecture that minimizes the loss function for 2000 epochs. For a comprehensive overview of the final optimized architecture, including its specifications, refer to Table~\ref{tab:optarch} in \ref{app:optarch}.

\subsection{Dimensionality Reduction/Expansion}
\begin{figure}[!b]
    \centering
    \includegraphics[width=1\linewidth]{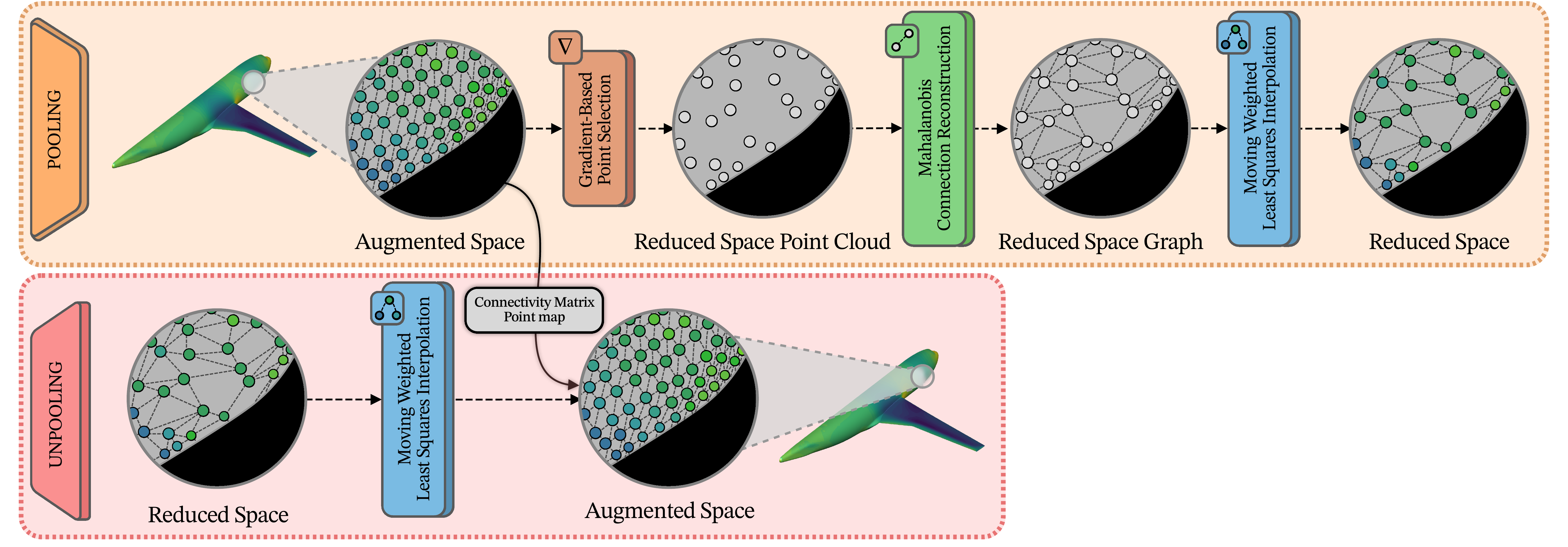}
    \caption{Pooling and unpooling modules architecture.}
    \label{fig:dimredexp}
\end{figure}

The core idea behind the use of space reduction/expansion operations is to minimize non--influential information from nodes that do not contribute to the nonlinearity of the system. The aim is to streamline the complexity of hidden layer operations and eliminate redundant information that could potentially mislead the model. The pooling and unpooling modules entail different concepts, which are herein explained. An overview is presented in Figure~\ref{fig:dimredexp}, where it is possible to distinguish all the processes used for construct and reconstruct hidden spaces. During encoding, we select points based on pressure gradients, creating a reduced--point cloud. We then use a Mahalanobis distance--based method to reconstruct connectivity, resulting in a connected graph. Node values are computed through grid interpolation using the moving weighted least squares method. In decoding, we interpolate on the original fine point map and connectivity using the same method with a new interpolation matrix.



\subsubsection*{Pressure gradient--based Point Selection}

The goal of gradient--based point selection is to find the optimal approach for implementing a pooling phase.  During this phase, points are chosen for removal from the mesh graph in the space reduction operation. The general idea is to employ a more advanced point selection method instead of relying solely on the simplistic density--based approach~\cite{massegur2023graph}. By doing so, the pooling phase can more effectively consider the primary region where nonlinear phenomena occur.

This method entails two fundamental steps. Initially, gradients on pressure value are computed for each sample. Then, the value of gradient for each example is used for the identification of regions of interest across the entire dataset. This approach facilitates the detection of areas where pressure gradients display notable differences on pressure, thereby identifying points characterized by heightened nonlinearity.

Spatial gradients are computed for each point by considering the pressure value at each node of the graph. To calculate gradients in unstructured grids, it is assumed that the pressure variable varies linearly in all dimensions, yielding:

\begin{equation}
    p - p_0 = \Delta p = \Delta x p_x + \Delta y p_y + \Delta z p_z
\end{equation}

Where $p_0$ is the pressure in the node. Then, a matrix equation is constructed using the pressure differences among all nodes neighboring the current node. With five connections, the matrix equation results in:

\begin{equation}
\begin{bmatrix}
\Delta x_1 & \Delta y_1 & \Delta z_1 \\
\Delta x_2 & \Delta y_2 & \Delta z_2 \\
\Delta x_3 & \Delta y_3 & \Delta z_3 \\
\Delta x_4 & \Delta y_4 & \Delta z_4 \\
\Delta x_5 & \Delta y_5 & \Delta z_5 
\end{bmatrix}
\begin{bmatrix}
p_x \\
p_y \\
p_z 
\end{bmatrix}
=
\begin{bmatrix}
\Delta p_1 \\
\Delta p_2 \\
\Delta p_3 \\
\Delta p_4 \\
\Delta p_5 
\end{bmatrix}\label{eq:grad}
\end{equation}

Equation~\eqref{eq:grad} is then inverted via the least-squares method to compute the gradient vector.

Starting form the value of the gradients calculated for all the points in the graph, a suitable probability distribution has been employed to determine the number of points retained in the reduced space. The challenge arises in regions of the original mesh with low gradients, potentially resulting in an inadequate number of nodes at the coarsened level and leading to an irreversible loss of information. Conversely, excessive node removal in regions of originally high gradients may result in insufficient accuracy reconstruction of complex physics phenomena. Thus, an appropriate node selection strategy is essential to ensure the proper representation of both high and low gradient regions in the coarsened domain. This is obtained using a probability function based on the gradient of the mesh element:

\begin{equation}
    p(i) = 1 + \frac{1 - e^{-2i/n}}{1 - e^{-2}} (p_1 - p_n) + p_1 \quad \text{for} \quad i = 1, \ldots, n
\end{equation}

Here, \(i\) represents the mesh node index, sorted by pressure gradient value in descending order, and \(n\) is the total number of nodes. The probabilities \(p_1\) and \(p_n\) denote the choices for the highest and lowest gradients, respectively set to 0.2 and 1.

After each space reduction, an unconnected point cloud is obtained, therefore it is essential to restore the connectivity between neighbours.

\subsubsection*{Mahalanobis connection reconstruction}

To identify the neighbors of each node in the point cloud after the reduction process and thereby restore connectivity, we use a reconstruction method based on the Mahalanobis distance~\cite{de2000mahalanobis}, that is widely used in clustering problems and other statistical classification techniques~\cite{xiang2008learning,ghorbani2019mahalanobis}. The Mahalanobis distance is a measure of the distance between points in a distribution. Unlike the simple Euclidean distance, the Mahalanobis distance takes into account the spread of points in different directions through the covariance matrix of the distribution of points. Using this type of distance, it is possible to connect each point to its neighbours by following the distribution of points in the finer mesh by using the covariance matrix calculated in the original space. This method minimizes false connections between opposite faces of the mesh which are considered close according to the simple Euclidean distance. Therefore, the distance between points is calculated using the following equation:

\begin{equation}
    D_M(x,y)=\sqrt{(x-y)^T S^{-1}(x-y)}
\end{equation}
Where $x$ and $y$ are two points of the reduced space and $S$ is the covariance matrix of the distribution of the points in the finer mesh. 

Additionally, to reduce the searching field of nearest neighbours on the reduced space, we used the K-d tree algorithm \cite{friedman1977algorithm} to determine  for each point a subset of 250 elements using Euclideian distance, and then selected the nearest neighbours by following the Mahalanobis distance calculated only in that subset.
\subsubsection*{Moving Weighted Least Squares for Grid Interpolation} \label{subsec:MWLS}

Efficient information transfer between grids is a critical aspect in the proposed methodology. While one option involves using a neural network with learnable weights, this approach could significantly escalate computational requirements. On the contrary, traditional interpolation techniques may yield inaccuracies that are not suitable for our purposes~\cite{massegur2023graph}. Consequently, we opted for the Moving Weighted Least Squares (MWLS) technique~\cite{quaranta2005conservative,joldes2015modified}. This decision aims to strike a balance between accuracy and computational efficiency, while also ensuring the conservation of the integrated quantity across both grids and maintaining continuity across the domain~\cite{quaranta2005conservative}. MWLS assigns varying weights to neighboring data points based on their proximity to the interpolation point, allowing for a more adaptive and accurate representation of the underlying data. The approach involves fitting a local polynomial to a subset of nearby points, with the influence of each point weighted according to its distance. This adaptability ensures that closer points have a more significant impact on the interpolated value, while those farther away contribute less.

The main idea is to generate an interpolation matrix \(I_{S_s \rightarrow S_d}\) that maps a feature \(\mathbf{y}_i\) from a source grid \(S_s\) containing \(n_s\) nodes to an interpolated solution feature \(\mathbf{y}_j\) on the destination grid \(S_d\) with \(n_d\) nodes, having both grids lying on the same spatial domain:

\begin{equation}
    \mathbf{y}_j = I_{S_s \rightarrow S_d} \mathbf{y}_i \quad \forall j \in S_d, \quad \forall i \in S_s 
\end{equation}


To accomplish this, a shape function \(u(\mathbf{x})\) approximating the grid data \(y_i\) evaluated at source nodes \(i \in S_s\) with coordinates \(\mathbf{x}_i\) must be generated by minimizing the least square error evaluated at these points:

\begin{equation}
    \min L = \sum_{i \in S_s} \left(u(\mathbf{x}_i) - y_i\right)^2 w(\mathbf{x}_i)
\end{equation}

The term $w(\mathbf{x}_i)$ represents the Gaussian weight function defined as: $w(\mathbf{x}_i) = e^{-{\|\mathbf{x} - \mathbf{x}_i\|_2}}$. This function is employed to assign higher weights to source nodes that are in close proximity to the destination node.

To construct \(u(\mathbf{x})\), a polynomial combination is adopted:
\begin{equation}
    u(\mathbf{x}) = \mathbf{p}^T(\mathbf{x}) \, \mathbf{a} 
\end{equation}

where \(\mathbf{p}(\mathbf{x})\) is a second--order polynomial basis function, i.e., \(\mathbf{p}(\mathbf{x}) = [1, x, y, z, x^2, y^2, z^2, xy, xz, yz]^T\), and \(\mathbf{a}\) is the vector of respective coefficients. 

The approximated value $u(\mathbf{x}_j)$ at every node destination node $j$ in $S_d$ can be obtained from the analytical solution of the least square minimization:

\begin{equation}
    u(\mathbf{x}_j) = \boldsymbol{\Phi}(\mathbf{x}_j) \mathbf{y}_{S_s} 
\end{equation}

The coefficients $\boldsymbol{\Phi}(\mathbf{x}_j)$ for each destination node are calculated as:

\begin{equation}
\label{eq:least_square_solution}
    \boldsymbol{\Phi}(\mathbf{x}_j) = \mathbf{p}^T(\mathbf{x}_j) (\mathbf{P}^T\mathbf{W}\mathbf{P})^{-1}\mathbf{P}^T\mathbf{W} 
\end{equation}

where $\mathbf{p}^T(\mathbf{x}_j)$ represents the polynomial basis for the destination node. The design matrix $\mathbf{P}$ is formed for the $n_s$ source nodes in $S_s$. The weight matrix $\mathbf{W}$ is constructed as a diagonal matrix with the Gaussian weights. 

\begin{equation}
\mathbf{P} =
\begin{bmatrix}
    \mathbf{p}^T(\mathbf{x}_1) \\
    \mathbf{p}^T(\mathbf{x}_2) \\
    \vdots \\
    \mathbf{p}^T(\mathbf{x}_n) \\
\end{bmatrix}
\end{equation}

\begin{equation}
    \mathbf{W} = \begin{bmatrix}
        w(\mathbf{x}_1)& 0 & \dots & 0 \\
        0 & w(\mathbf{x}_2) & \dots & 0 \\
        \vdots & \vdots & \ddots & \vdots \\
        0 & 0 & \dots & w(\mathbf{x}_{n_s})
    \end{bmatrix} 
\end{equation}

The coefficients $\boldsymbol{\Phi}(\mathbf{x}_j)$ for each of the $n_d$ destination nodes are stored in the interpolation matrix \(I_{S_s \rightarrow S_d}\) which is then used in the pooling layer:

\begin{equation}
    I_{S_s \rightarrow S_d} = \begin{bmatrix}
        \boldsymbol{\Phi}(\mathbf{x}_1) \\
        \boldsymbol{\Phi}(\mathbf{x}_2) \\
        \vdots \\
        \boldsymbol{\Phi}(\mathbf{x}_{n_d})
    \end{bmatrix} 
\end{equation}

The computation of the least squares solution, Equation~\eqref{eq:least_square_solution}, is required for each node of the destination grid. To reduce the burden of such computation, a local (i.e., moving) interpolation is adopted by imposing that each destination node is only influenced by the \(k_n\) closest source nodes:

\begin{equation}
    w(\mathbf{x}_i) = \begin{cases}
        e^{-{\|\mathbf{x} - \mathbf{x}_i\|_2}} & \text{for the } k_n \text{ nearest source nodes in } S_s \\
        0 & \text{for the remaining nodes}
    \end{cases}  
\end{equation}

The optimal number of neighbors was found to be \(k_n=10\), striking a balance between minimizing reconstruction errors and managing computational requirements efficiently. 

It is worth remarking that this interpolated matrix is of non-square size \(n_s \times n_d\) and largely sparse, with only \(k_n\) non-zero values in each row. Consequently, with regards to executing the inverse interpolation in the decoder phase, this matrix is not invertible. Thus, it is necessary to compute two independent interpolation matrices: \(I_{S_s \rightarrow S_d}\) and \(I_{S_d \rightarrow S_s}\). 

\section{Test Cases}\label{sec:results}

\begin{figure} [!htb] 
    \centering
    \includegraphics[trim=1 1 1 1, clip, width=0.88\textwidth]{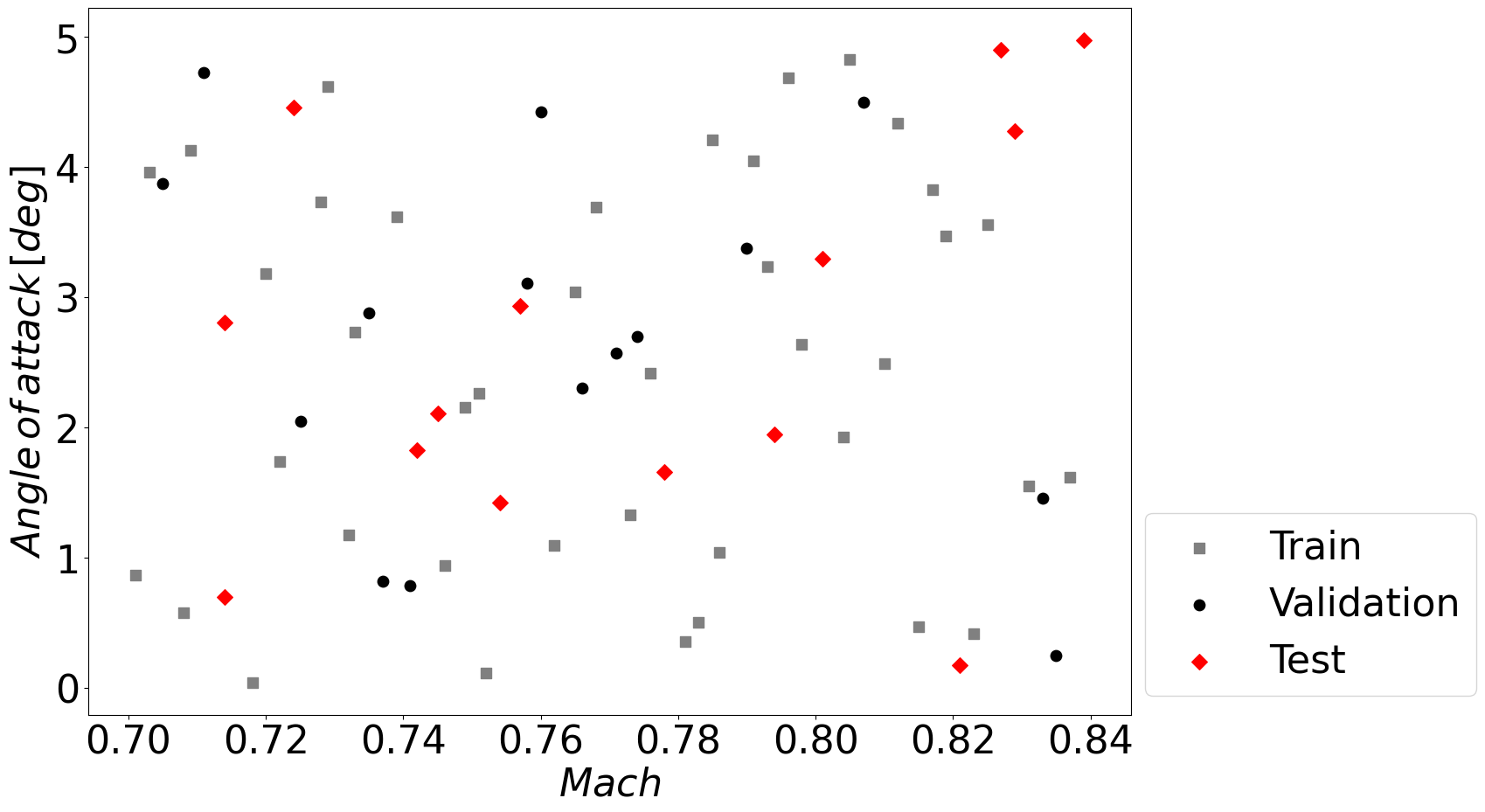}
    \caption{Training, validation and test samples for Mach number and angle of attack.}
    \label{fig_sampled_points}
\end{figure}

Two test cases, characterized by different physics and complexity, were employed for assessing the model prediction capability. Angle of attack and  Mach number were chosen as the two independent parameters for the ROM. The chosen ranges for the angle of attack ($AoA$) and Mach number ($M$) are $[0, 5] \, [deg]$ and $[0.70, 0.84]$, respectively. These ranges are specifically chosen for the transonic regime, where shock wave formation occurs on the wing, and high angles of attack, that lead to boundary--layer separation. To generate the required number of samples, Latin hypercube sampling (LHS) \cite{loh1996latin} is employed, resulting in a total of 70 points as illustrated in Figure~\ref{fig_sampled_points}. Sixty percent of these samples (40 flight conditions denoted by circles) are designated for training, $20\%$ (15 flight conditions marked with squares) for validation, and the remaining $20\%$ (15 conditions represented by diamonds) are reserved for testing.

The dataset has been generated through CFD simulations. Reynolds-averaged Navier–Stokes (RANS) equations are discretized using SU2 v7.5.1~\cite{SU2} software. The closure of RANS equations is achieved using the one--equation Spalart--Allmaras turbulence model. Convergence method is set to Cauchy method, specifically applied to the lift coefficient, considering a variation of $10^{-7}$ across the last $100$ iterations. A $1v$ multigrid scheme is adopted for accelerating the convergence of CFD simulations. The discretization of convective flows involves the use of the Jameson-Schmidt-Turkel (JST) central scheme with artificial dissipation. Flow variable gradients are computed through the Green Gauss method. The selected linear solver is the biconjugate gradient stabilization, with an ILU preconditioner.

Following dataset generation, preprocessing and normalization to the range $[-1,1]$ were performed before inputting the data into the AE--GCN model. The rest of the section explores the model predictive capabilities for distributed quantities and integral loads across different test cases. Computational performance and optimized architectures are detailed in \ref{app:optarch}.

\subsection{Wing--only Model}

The first test case is the Benchmark Super Critical Wing (BSCW), which is a transonic rigid semi--span wing with a rectangular planform and a supercritical airfoil shape from the AIAA Aeroelastic Prediction Workshop~\cite{heeg2013overview}. This wing is elastically suspended on a flexible mount system with two degrees of freedom, pitch and plunge, and it has been developed for flutter analysis as it is characterized by shock wave motion, shock--induced boundary--layer separation and interaction between shock wave and detached boundary--layer. These three types of nonlinearity are challenging for the ROM predictions. 

\begin{figure} [!htb] 
    \centering
    \includegraphics[trim=1 1 1 1, clip, width=0.68\textwidth]{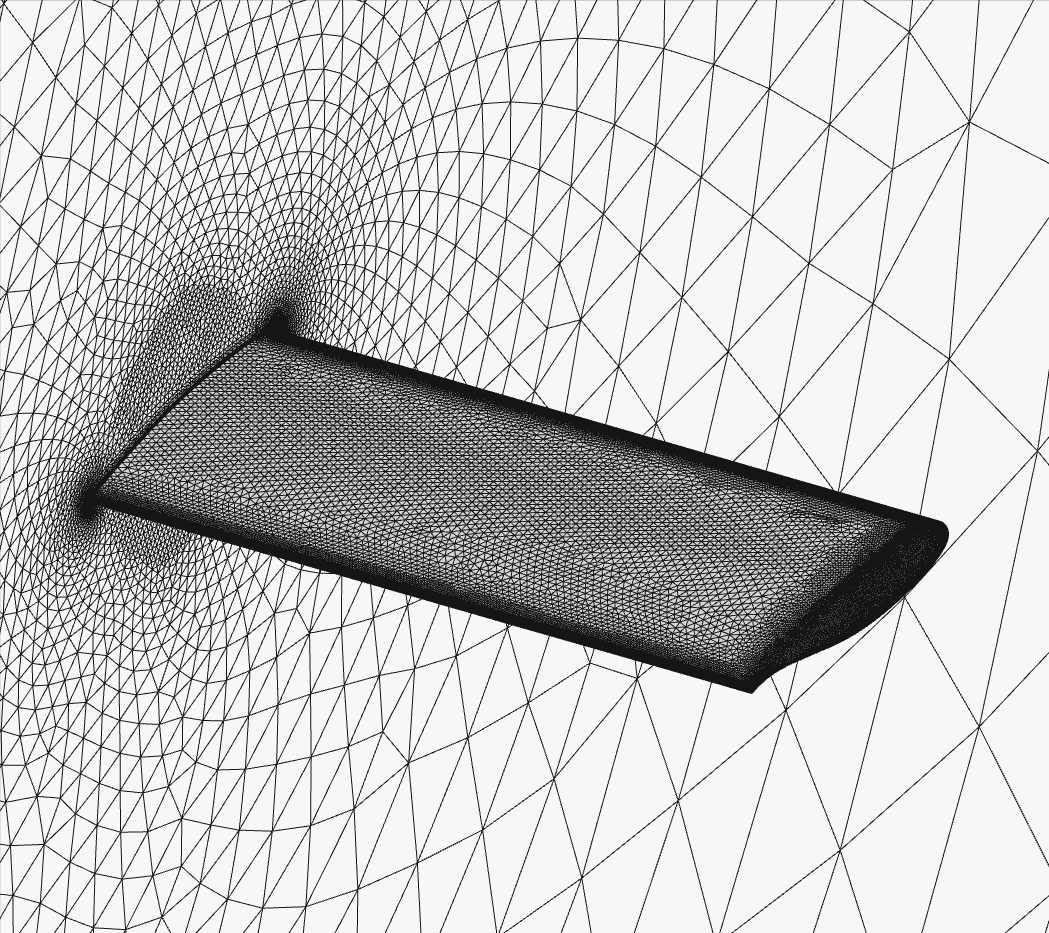}
    \caption{Impression of the BSCW CFD grid.}
    \label{fig_bscw_mesh}
\end{figure}

An unstructured grid configuration with $8.4 \cdot10^6$ elements and 86,840 surface elements was generated. A $y^+ = 1$ is adopted, after a preliminary mesh convergence study that ensured an adequate resolution of the boundary--layer and shock wave. The computational domain extends 100 chords from the solid wall to the farfield. An impression of the grid can be obtained from Figure~\ref{fig_bscw_mesh}.

The results obtained with the implemented model are herein presented. The Mean Absolute Percentage Error (\texttt{MAPE}) was computed by averaging the absolute error of each prediction calculated by our AE-GCN architecture within the test set. This prediction error was determined by weighted averaging the errors at each grid point, considering the corresponding cell area and normalizing with respect to it. The results reveal a particularly low \texttt{MAPE} values of 0.7712 for $C_P$ and 0.3828 for $C_F$.

\begin{figure} [!htb] 
    \centering
    \includegraphics[trim=0.1cm 0.1cm 0.1cm 0.1cm , clip, width=0.98\textwidth]{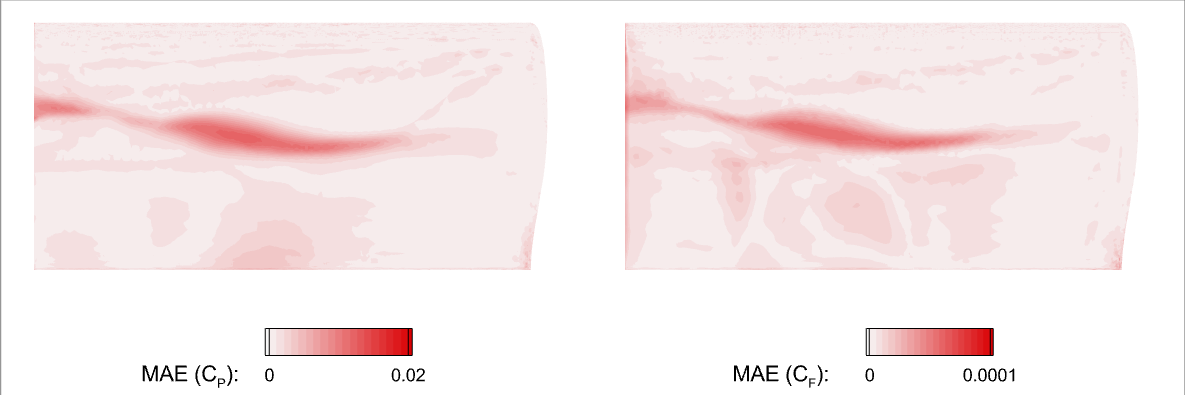} \\
    \caption{Mean absolute error (MAE) of $C_P$ and $C_F$ computed across every point in the mesh of the test set for BSCW test case.}
    \label{fig_mae_cp_cf_surface_BSCW}
\end{figure}

Figure~\ref{fig_mae_cp_cf_surface_BSCW} depicts the mean absolute error (\texttt{MAE}) of $C_P$ and $C_F$ calculated across each point in the test set mesh for the BSCW model. Remarkably, the errors for both predictions are considerably small, with the selected ranges serving solely to offer a visual depiction of areas where the model faces challenges in prediction. The errors are minimal across the entire surface, except for a localized region near the shock wave.

Figure~\ref{fig_errors_sampled_points_crm} illustrates the percentage errors in $[C_L,C_D,C_{My}]$ across different Mach numbers and angles of attacks for the BSCW test case. Remarkably, the model predictions exhibit high accuracy for all coefficients, even for data points located far from the training set. This indicates the model robustness in extrapolating beyond the provided data points. Aerodynamic coefficients were calculated using a reference chord length of 0.4064 $m$ and surface of 0.3303 $m^2$, and derived by integrating the pressure coefficient distribution and the skin friction coefficient distribution over the entire wing surface. $C_{My}$ was calculated with respect to $30\%$ of the chord, accounting for the rigid mounting system of the BSCW, which induces pitch oscillations around this specific location. 

\begin{figure} [!htb] 
    \centering
    \includegraphics[trim=1 1 1 1, clip, width=0.98\textwidth]{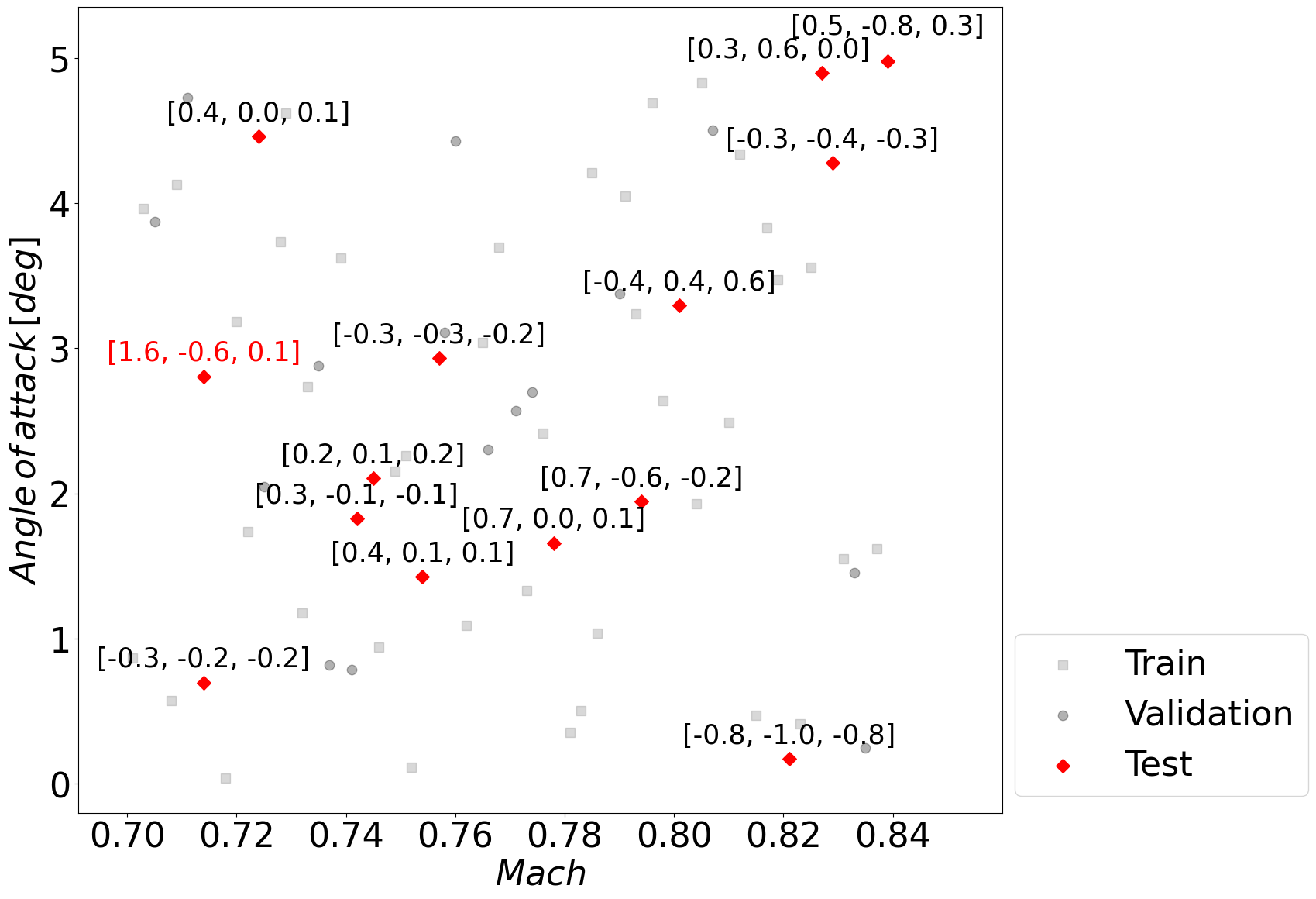}
    \caption{Errors $\%$ in $[C_L,C_D,C_{My}]$ on the test samples across varying Mach numbers and angle of attacks for BSCW test case.}
    \label{fig_errors_sampled_points_BSCW}
\end{figure}

\begin{figure} [!htb] 
    \centering
    \includegraphics[trim=0.1cm 0.1cm 0.1cm 0.1cm , clip, width=0.98\textwidth]{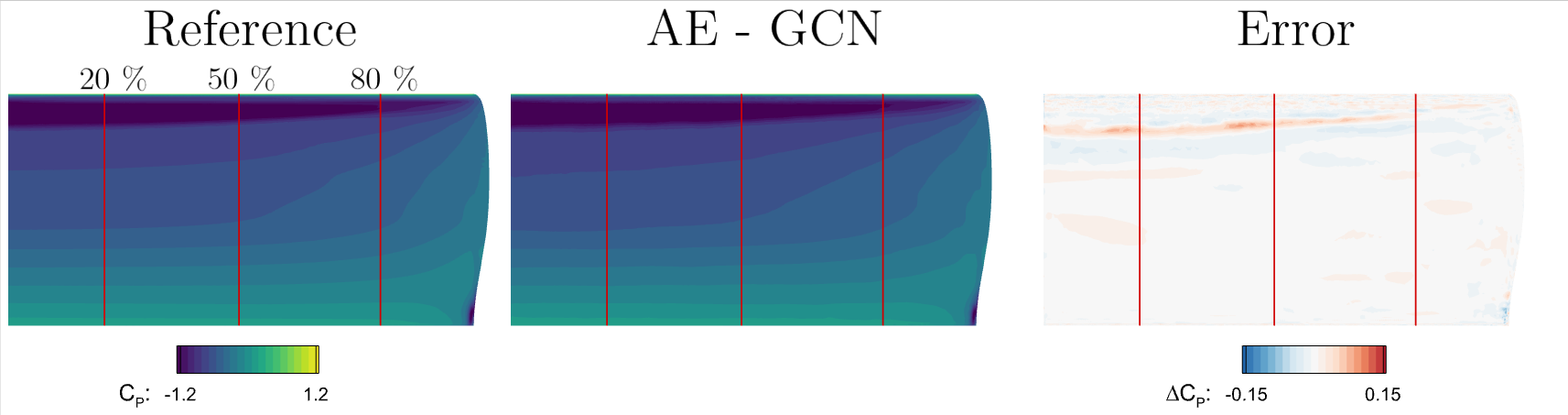} \\
    \caption{Prediction of the pressure coefficient contour on the upper surface for BSCW test case at $M = 0.714$ and $AoA = 2.807 \,\, [deg]$.}
    \label{fig_cp_surface_bscw}
\end{figure}

\begin{figure} [!htb] 
    \centering
    \includegraphics[trim=0.1cm 0.1cm 0.1cm 0.1cm , clip, width=0.98\textwidth]{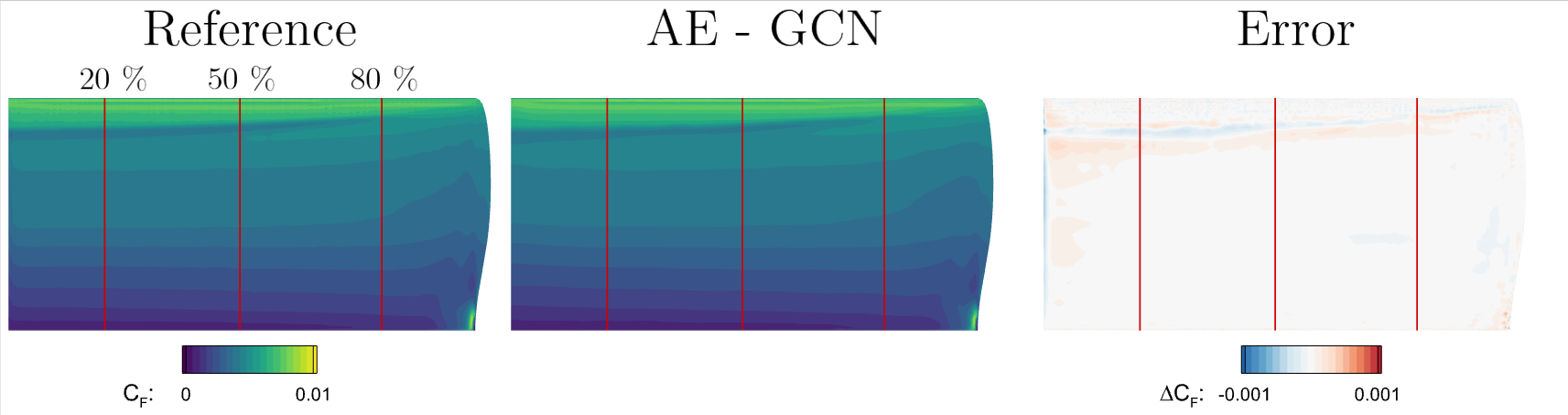} \\
    \caption{Prediction of the skin friction magnitude coefficient contour on the upper surface for BSCW test case at $M = 0.714$ and $AoA = 2.807 \,\, [deg]$.}
    \label{fig_cf_surface_bscw}
\end{figure}

Figure~\ref{fig_cp_surface_bscw} displays a comparison between the pressure coefficient contour of CFD data and the reconstructed surface field with AE--GCN model. This comparison is made for the test sample with the highest error at $M = 0.714$ and $AoA = 2.807 \,\, [deg]$, with the wing root positioned on the left side and the incoming flow directed onto the wing from the leading edge. A remarkable agreement is observed between CFD and the ROM, especially in predicting the strong nonlinear pressure distribution in terms of shock wave position and size. A small error might be noticed in correspondence of the low pressure area near the shock wave. Similar observations can be made when analyzing the skin friction contour in Figure~\ref{fig_cf_surface_bscw}. 

\begin{figure} [!htb] 
    \centering
    \includegraphics[trim=5 5 5 5, clip, width=0.98\textwidth]{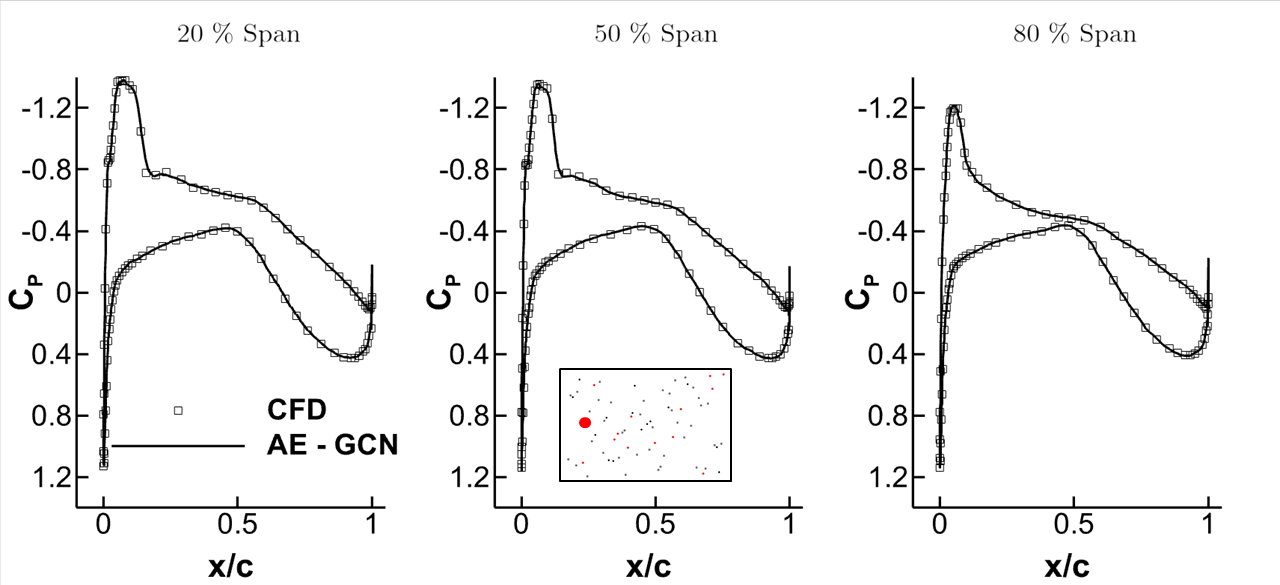} 
    \caption{Pressure coefficient sections of BSCW at $M = 0.714$ and $AoA = 2.807 \,\, [deg]$.}
    \label{fig_cp_sections_bscw}
\end{figure}

\begin{figure} [!htb] 
    \centering
    \includegraphics[trim=5 5 5 5, clip, width=0.98\textwidth]{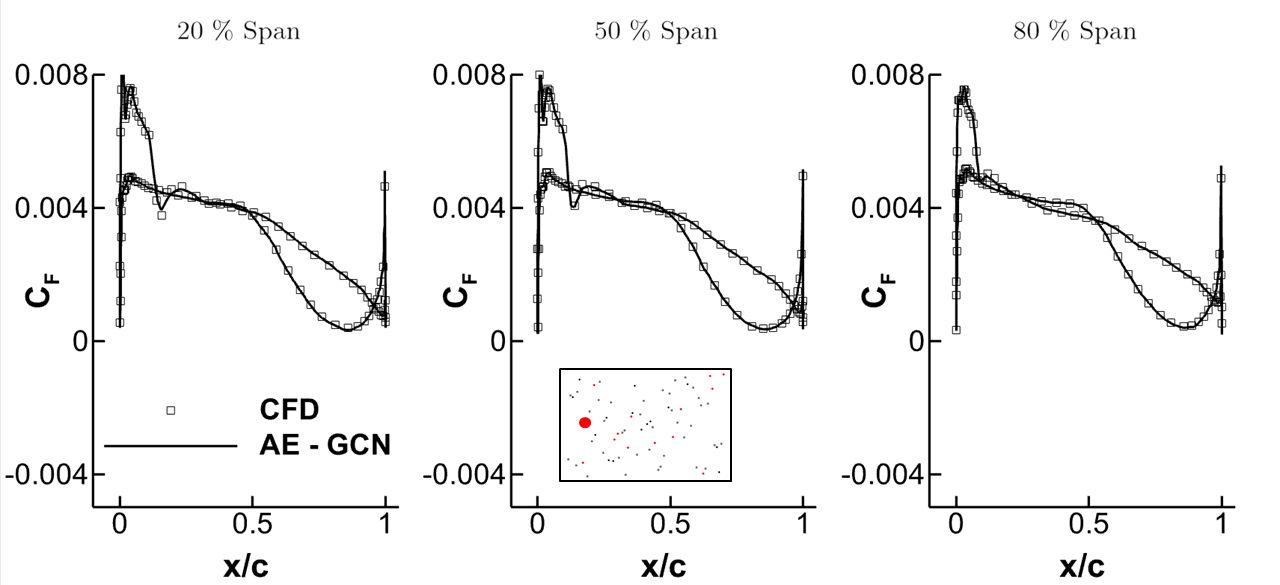} 
    \caption{Skin friction coefficient sections of BSCW at $M = 0.714$ and $AoA = 2.807 \,\, [deg]$.}
    \label{fig_cf_sections_bscw}
\end{figure}

Figures~\ref{fig_cp_sections_bscw} and \ref{fig_cf_sections_bscw} provide detailed view of the pressure and skin friction coefficient distribution at three distinct sections along the span of the BSCW test case, evaluated at $M = 0.714$ and $AoA = 2.807 \,\, [deg]$. Notably, the ROM exhibits precise predictions of pressure peaks, particularly in the vicinity of the shock location occurring at $20\%$ of the wing span. This highlights the model capability to capture critical aerodynamic features with a high degree of fidelity.

\subsection{Wing--body Model}

\begin{figure} [!htb] 
    \centering
    \includegraphics[trim=1 1 1 1, clip, width=0.68\textwidth]{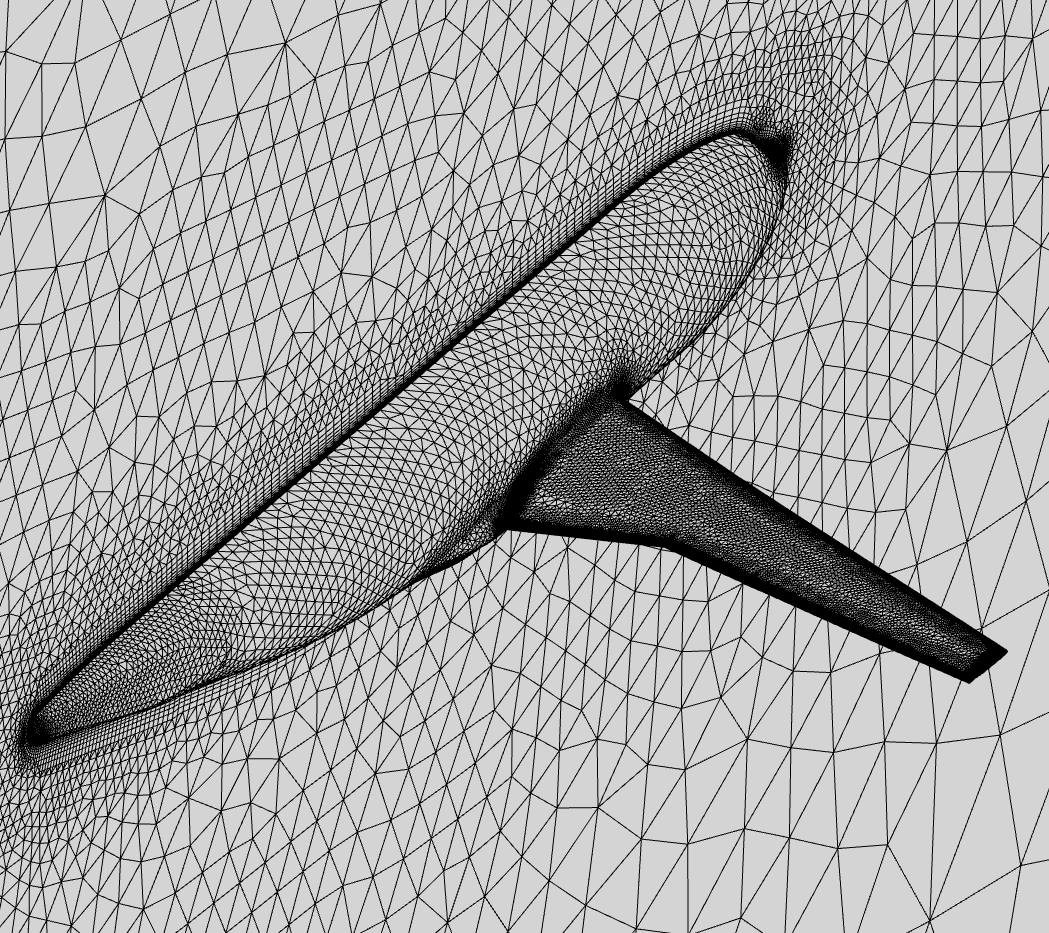}
    \caption{Impression of the CRM CFD grid.}
    \label{fig_crm_mesh}
\end{figure}

The second test case is the NASA Common Research model (CRM), a transonic wing--body model featured in the AIAA CFD Drag Prediction Workshop~\cite{vassberg2008abridged}. This model encompasses a conventional low--wing configuration and a fuselage typical of wide--body commercial aircraft. The computational grid utilized for this case was adapted from the DLR grid developed for the AIAA Drag Prediction Workshop~\cite{vassberg2007summary}. This unstructured grid comprises $8.8 \times 10^6$ elements, including 78,829 surface elements. The computational domain extends 100 chords from the fuselage to the farfield. A $y^+ = 1$ condition is employed. For a visual representation of the grid, refer to Figure~\ref{fig_crm_mesh}.

This test case poses a complex challenge for our AE--GCN model due to the complex geometry, physics and grid configuration. It differs from the previous one as it has a remarkable amount of surface points in areas characterized by predominantly linear flow, such as vast regions of the fuselage. Consequently, we opted to evaluate the \texttt{MAPE} exclusively on the wing, where nonlinearity is pronounced. The results indicate a \texttt{MAPE} of 0.8876 for $C_P$ and 0.2402 for $C_F$.

\begin{figure} [!htb] 
    \centering
    \includegraphics[trim=0.1cm 0.1cm 0.1cm 0.1cm , clip, width=0.98\textwidth]{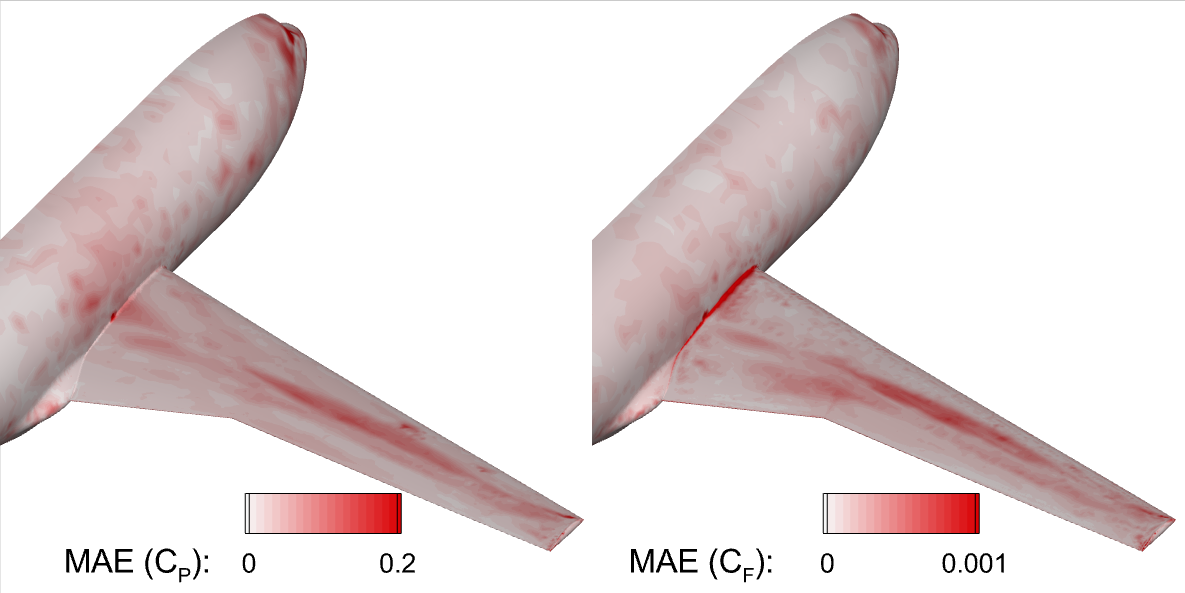} \\
    \caption{Mean absolute error (MAE) of $C_P$ and $C_F$ computed across every point in the mesh of the test set for CRM test case.}
    \label{fig_mae_cp_cf_surface_crm}
\end{figure}

The mean absolute error (\texttt{MAE}) of $C_P$ and $C_F$ computed across every point in the mesh of the test set is depicted in Figure~\ref{fig_mae_cp_cf_surface_crm}. This visualization provides insight into the regions where the model struggles most to accurately represent the flow physical behavior. Interestingly, the errors are generally minimal across the entire surface, except for a localized region near the wing-fuselage junction and between the kink of the wing and its tip. Nonetheless, the broadly distributed small errors suggest that the model effectively captures the nonlinearities inherent in the system.

\begin{figure} [!htb] 
    \centering
    \includegraphics[trim=1 1 1 1, clip, width=0.98\textwidth]{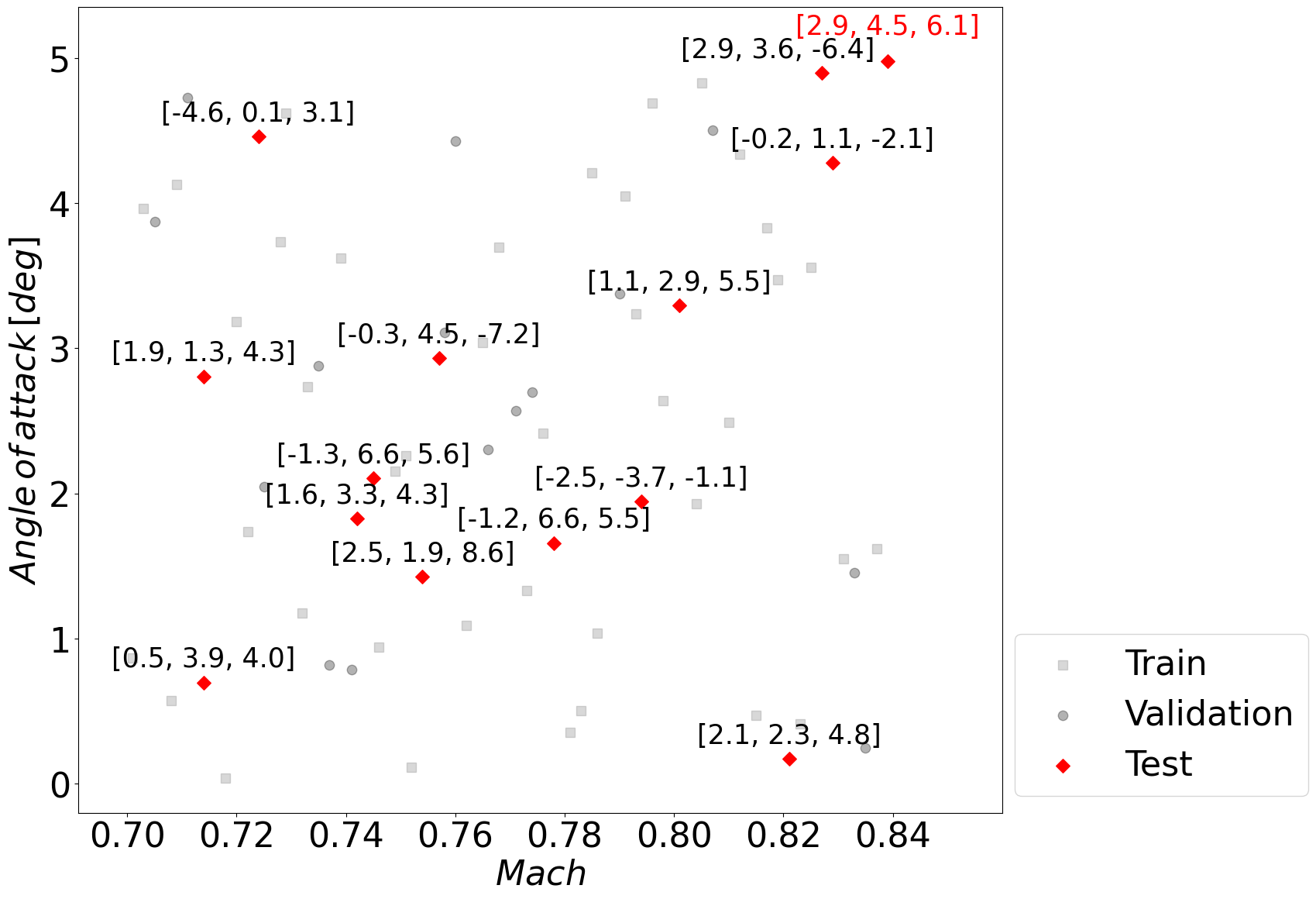}
    \caption{Errors $\%$ in $[C_L,C_D,C_{My}]$ on the test samples across varying Mach numbers and angle of attacks for CRM test case.}
    \label{fig_errors_sampled_points_crm}
\end{figure}

Figure~\ref{fig_errors_sampled_points_crm} shows the percentage errors in $[C_L,C_D,C_{My}]$ on the test samples across varying Mach numbers and angle of attacks. Notably, the predictions demonstrate overall accuracy across all coefficients, even for points distant from the training samples. This suggests a robust performance of the model in extrapolating beyond the known data points. A chord of 0.1412 $m$ and a scaling area of 0.1266 $m^2$ were considered for aerodynamic coefficients calculation. $C_{My}$ was computed with respect to $25\%$ of the the wing mean aerodynamic chord.

\begin{figure} [!htb] 
    \centering
    \includegraphics[trim=0.1cm 0.1cm 0.1cm 0.1cm , clip, width=0.98\textwidth]{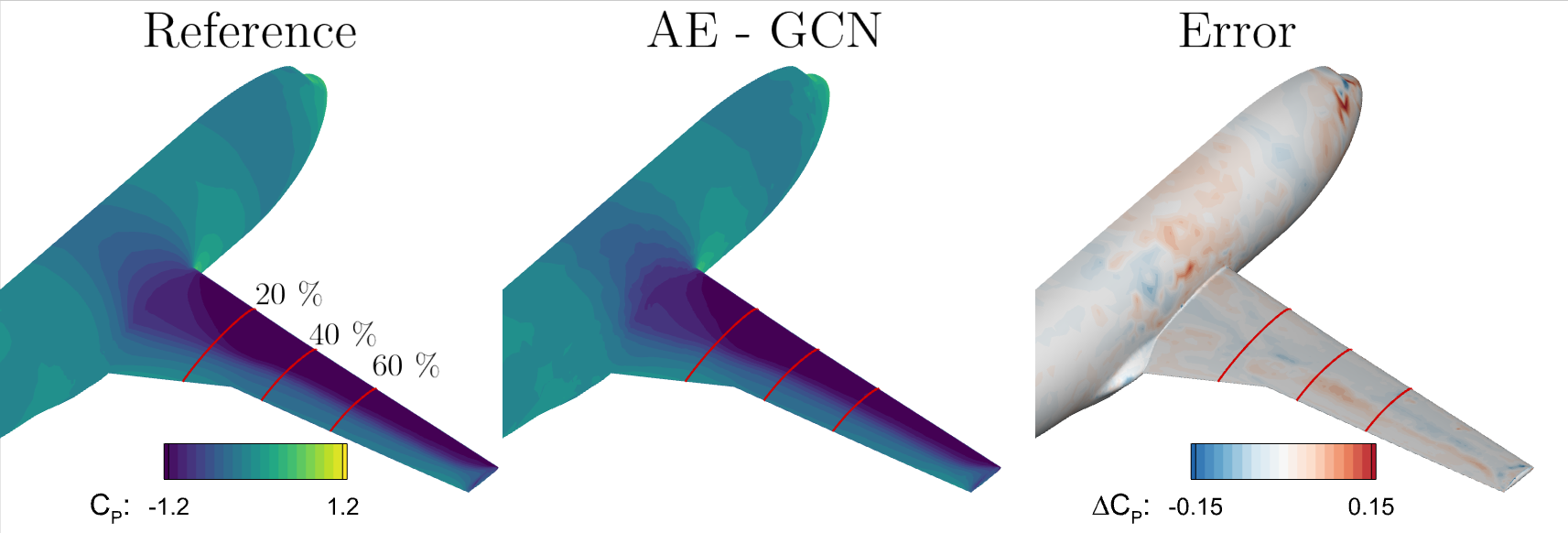} \\
    \caption{Prediction of the pressure coefficient contour on the upper surface for CRM test case at $M = 0.839$ and $AoA = 4.975 \,\, [deg]$.}
    \label{fig_cp_surface_crm}
\end{figure}

\begin{figure} [!htb] 
    \centering
    \includegraphics[trim=0.1cm 0.1cm 0.1cm 0.1cm , clip, width=0.98\textwidth]{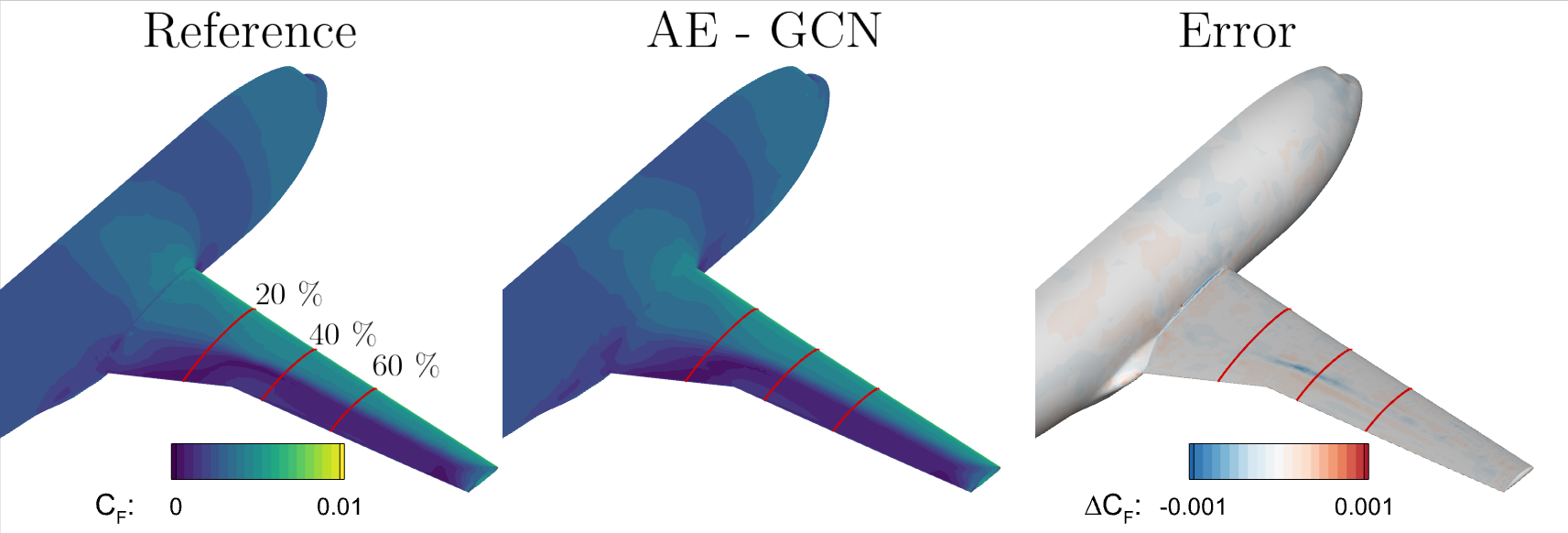} \\
    \caption{Prediction of the skin friction magnitude coefficient contour on the upper surface for CRM test case at $M = 0.839$ and $AoA = 4.975 \,\, [deg]$.}
    \label{fig_cf_surface_crm}
\end{figure}

Surface pressure contour and skin friction contour predictions are shown respectively in Figures~\ref{fig_cp_surface_crm} and \ref{fig_cf_surface_crm} on the worst prediction of the test sample at $M = 0.839$ and $AoA = 4.975 \,\, [deg]$. A good alignment with the reference CFD data is evident, indicating a favorable agreement. Notably, the error is distributed throughout the entire aircraft, indicating that the model effectively captures the underlying physical phenomena occurring under these specific flight conditions. Moreover, this highlights the model ability to generalize and obtain accurate predictions even in challenging scenarios.

\begin{figure} [!htb] 
    \centering
    \includegraphics[trim=5 5 5 5, clip, width=0.98\textwidth]{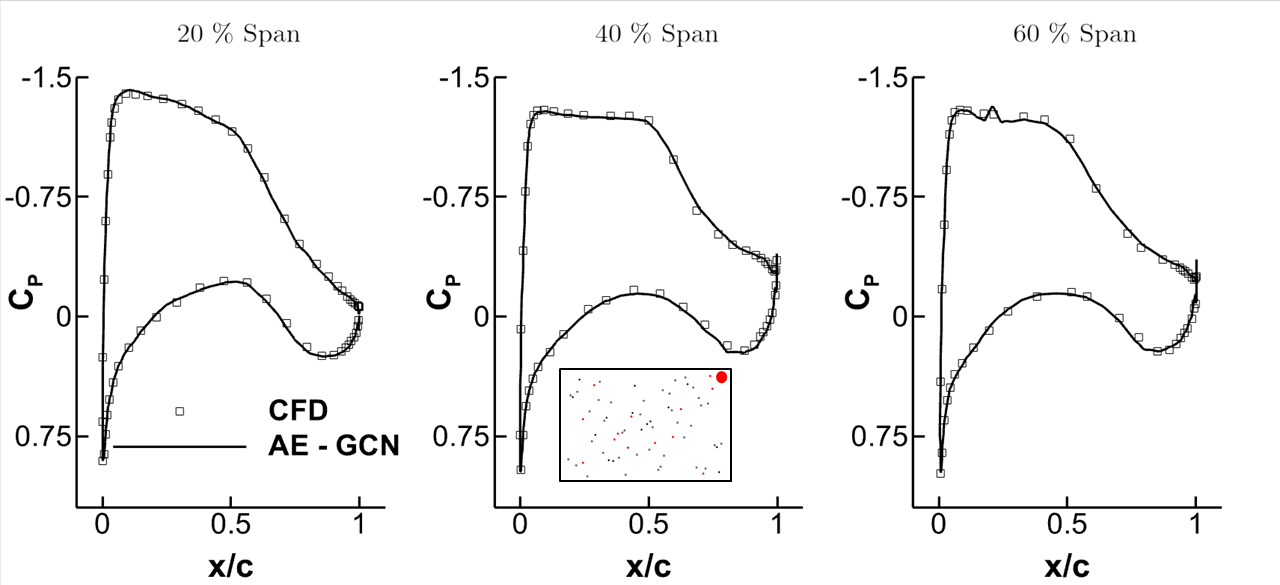} 
    \caption{Pressure coefficient sections of CRM at $M = 0.839$ and $AoA = 4.975 \,\, [deg]$.}
    \label{fig_cp_sections_crm}
\end{figure}

\begin{figure} [!htb] 
    \centering
    \includegraphics[trim=5 5 5 5, clip, width=0.98\textwidth]{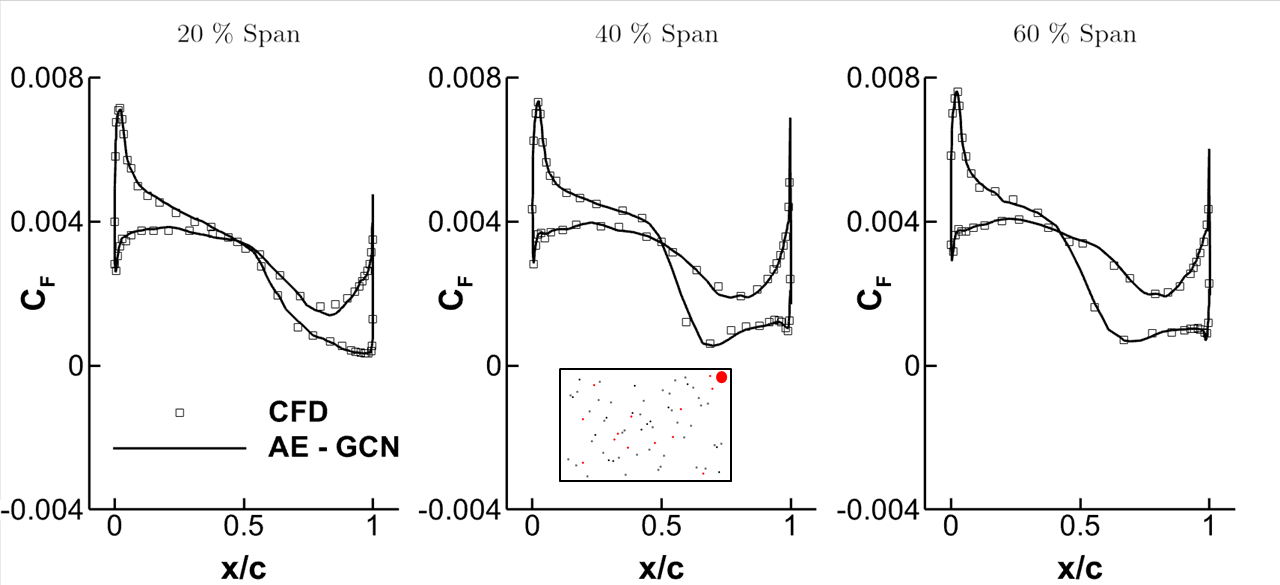} 
    \caption{Skin friction coefficient sections of CRM at $M = 0.839$ and $AoA = 4.975 \,\, [deg]$.}
    \label{fig_cf_sections_crm}
\end{figure}

Pressure coefficient prediction across several sections along the span is depicted in Figure~\ref{fig_cp_sections_crm}. Notably, there is a precise alignment with the reference data in each section. Similarly, the skin friction coefficient, highlighted in three sections along the span in Figure~\ref{fig_cf_sections_crm}, also demonstrates consistent agreement with the reference data.

The same network architecture was also tested using a point selection method based on the grid points density in the dimensionality reduction module. An observed error increase of approximately $3\%$ for $C_P$ and of $2\%$ for $C_F$ was noted on \texttt{MAPE} of test samples, particularly showing higher error in correspondence of nonlinear phenomena in terms of shock wave location and intensity. This outcome underscores the robustness of the proposed methodology for selecting the points to retain in the reduced space which permits to capture the higher nonlinearities.

\section{Conclusions}\label{sec:Conclusions}

Our study has demonstrated the effectiveness and robustness of the implemented model in delivering precise predictions within the parameter space. Through the utilization of convolutional and pooling operations, the model showcased its efficacy in influencing predictions of individual nodes based on their neighbors, while also facilitating information propagation to distant nodes during spatial reduction. A significant advantage of our model is its ability to directly process input grids without requiring preprocessing, simplifying the modeling process considerably.


Furthermore, our model exhibited high accuracy across various test scenarios, including those featuring complex geometries and diverse physical phenomena. This underscores its versatility and reliability in practical settings. The optimization of the network played a crucial role in achieving such accuracy, highlighting the importance of fine--tuning hyperparameters for each test case.

Additionally, the applicability of our model extends beyond the aerospace field, encompassing any non--homogeneous unstructured type data. This n

Looking ahead, a promising avenue for extending our current work involves exploring the modeling of unsteady--state phenomena. By incorporating temporal dynamics into our framework, we aim to enhance our model ability to capture transient behaviors and dynamic changes over time. This would further broaden its applicability and relevance in dynamical systems analysis.

\section*{Acknowledgement}

This work was supported by Digitalization Initiative of the Zurich Higher Education Institutions (DIZH) grant from Zurich University of Applied Sciences (ZHAW). The authors also acknowledge the University of Southampton for granting access to the IRIDIS High Performance Computing Facility and its associated support services.

\appendix

\section{Optimized AE--GCN Architecure} \label{app:optarch}

This section provides a comprehensive overview of the optimized architectures and highlights the systematic reduction of loss throughout the optimization trials for each test case.

Table~\ref{tab:optarch_bscw} provides detailed information about the optimized architecture designed specifically for the wing--only test case. This architecture consists of 17 layers and a total of 711,493 parameters, carefully balanced to capture the complexities of this aerodynamic setup. Similarly, Table~\ref{tab:optarch} displays the optimized architecture for the wing--fuselage test case. With 15 layers and a total of 633,731 parameters, this configuration is tailored to accurately model the interaction between the wing and fuselage, capturing the subtle aerodynamic interactions between these components.

\begin{table}[!t]
\small
\centering
\begin{tabular}{lllll}
\hline
\hline
                               & \textbf{Block}                  & \textbf{Layer} & \textbf{Activation} & \textbf{Output Size} \\ \hline \hline
Input                          &                                 &                &                     & m $\times$ 86840 $\times$ 5        \\ \hline
\multirow{7}{*}{Encoding}      & Block 0                         & GCN            & PReLU               & m $\times$ 86840 $\times$ 64      \\ \cline{2-5} 
                               & \multirow{3}{*}{Block 1}        & GCN            & PReLU               & m $\times$ 86840 $\times$ 112      \\
                               &                                 & GCN            & PReLU               & m $\times$ 86840 $\times$ 192      \\ 
                               &                                 & GCN            & PReLU               & m $\times$ 86840 $\times$ 256      \\ \cline{2-5} 
                               & Pooling 1                       &                &                     & m $\times$ 28600 $\times$ 256      \\ \cline{2-5} 
                               & \multirow{2}{*}{Block 2}        & GCN            & PReLU               & m $\times$ 28600 $\times$ 256      \\
                               &                                 & GCN            & PReLU               & m $\times$ 28600 $\times$ 288      \\ \cline{2-5} 
                               & Pooling 2                       &                &                     & m $\times$ 9600 $\times$ 288       \\ \hline
\multirow{2}{*}{Reduced Space} & \multirow{2}{*}{Block 3}        & GCN            & PReLU               & m $\times$ 9600 $\times$ 496       \\
                               &                                 & GCN            & PReLU               & m $\times$ 9600 $\times$ 288       \\ \hline
\multirow{6}{*}{Decoding}      & Unpooling 2              &                &                     & m $\times$ 28600 $\times$ 288      \\ \cline{2-5} 
                               & \multirow{2}{*}{Block 4} & GCN            & PReLU               & m $\times$ 28600 $\times$ 256      \\
                               &                          & GCN            & PReLU               & m $\times$ 28600 $\times$ 256      \\ \cline{2-5} 
                               & Unpooling 1              &                &                     & m $\times$ 28600 $\times$ 256      \\ \cline{2-5} 
                               & \multirow{3}{*}{Block 5} & GCN            & PReLU               & m $\times$ 86840 $\times$ 256      \\
                               &                          & GCN            & PReLU               & m $\times$ 86840 $\times$ 192      \\
                               &                          & GCN            & PReLU               & m $\times$ 86840 $\times$ 160      \\ \hline
\multirow{6}{*}{Output}        & \multirow{4}{*}{Block 6} & GCN            & PReLU               & m $\times$ 86840 $\times$ 1        \\
                               &                          & GCN            & PReLU               & m $\times$ 86840 $\times$ 1        \\
                               &                          & GCN            & PReLU               & m $\times$ 86840 $\times$ 1        \\
                               &                          & GCN            & PReLU               & m $\times$ 86840 $\times$ 1        \\ \cline{2-5} 
                               & \multicolumn{4}{c}{Concatenate Block 6}                                                \\ \cline{2-5} 
                               & Prediction               &                &                     & m $\times$ 86840 $\times$ 4        \\ \hline
\end{tabular}
\caption{Optimal architecture for Test Case I - Wing--only Model.}\label{tab:optarch_bscw}
\end{table}

\begin{table}[!htb]
\small
\centering
\begin{tabular}{lllll}
\hline
\hline
                               & \textbf{Block}                  & \textbf{Layer} & \textbf{Activation} & \textbf{Output Size} \\ \hline \hline
Input                          &                                 &                &                     & m $\times$ 78829 $\times$ 5        \\ \hline
\multirow{7}{*}{Encoding}      & Block 0                         & GCN            & PReLU               & m $\times$ 78829 $\times$ 224      \\ \cline{2-5} 
                               & \multirow{2}{*}{Block 1}        & GCN            & PReLU               & m $\times$ 78829 $\times$ 192      \\
                               &                                 & GCN            & PReLU               & m $\times$ 78829 $\times$ 192      \\ \cline{2-5} 
                               & Pooling 1                       &                &                     & m $\times$ 26000 $\times$ 192      \\ \cline{2-5} 
                               & \multirow{2}{*}{Block 2}        & GCN            & PReLU               & m $\times$ 26000 $\times$ 240      \\
                               &                                 & GCN            & PReLU               & m $\times$ 26000 $\times$ 304      \\ \cline{2-5} 
                               & Pooling 2                       &                &                     & m $\times$ 8000 $\times$ 304       \\ \hline
\multirow{2}{*}{Reduced Space} & \multirow{2}{*}{Block 3}        & GCN            & PReLU               & m $\times$ 8000 $\times$ 432       \\
                               &                                 & GCN            & PReLU               & m $\times$ 8000 $\times$ 304       \\ \hline
\multirow{6}{*}{Decoding}      & Unpooling 2              &                &                     & m $\times$ 26000 $\times$ 304      \\ \cline{2-5} 
                               & \multirow{2}{*}{Block 4} & GCN            & PReLU               & m $\times$ 26000 $\times$ 240      \\
                               &                          & GCN            & PReLU               & m $\times$ 26000 $\times$ 192      \\ \cline{2-5} 
                               & Unpooling 1              &                &                     & m $\times$ 26000 $\times$ 192      \\ \cline{2-5} 
                               & \multirow{2}{*}{Block 5} & GCN            & PReLU               & m $\times$ 78829 $\times$ 192      \\
                               &                          & GCN            & PReLU               & m $\times$ 78829 $\times$ 64      \\ \hline
\multirow{6}{*}{Output}        & \multirow{4}{*}{Block 6} & GCN            & PReLU               & m $\times$ 78829 $\times$ 1        \\
                               &                          & GCN            & PReLU               & m $\times$ 78829 $\times$ 1        \\
                               &                          & GCN            & PReLU               & m $\times$ 78829 $\times$ 1        \\
                               &                          & GCN            & PReLU               & m $\times$ 78829 $\times$ 1        \\ \cline{2-5} 
                               & \multicolumn{4}{c}{Concatenate Block 6}                                                \\ \cline{2-5} 
                               & Prediction               &                &                     & m $\times$ 78829 $\times$ 4        \\ \hline
\end{tabular}
\caption{Optimal architecture for Test Case II - Wing--fuselage Model.}\label{tab:optarch}
\end{table}

\begin{figure}[!hb]
\centering

\subfigure[Test Case I - Wing--only Model]{
\includegraphics[trim=0 0 0 0, clip, width=0.45\linewidth]{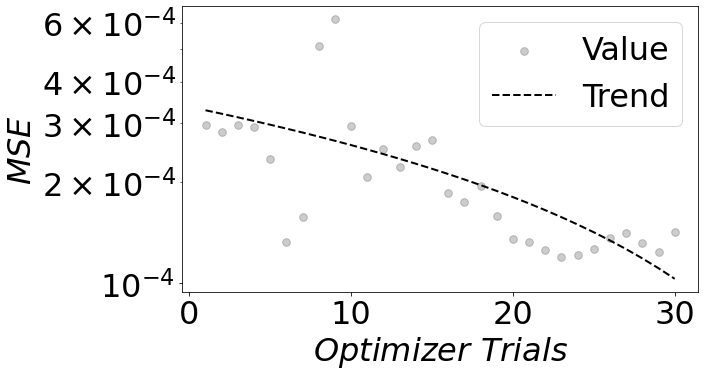}}
\hfill
\subfigure[Test Case II - Wing--fuselage Model]{
\includegraphics[trim=0 0 0 0, clip, width=0.45\linewidth]{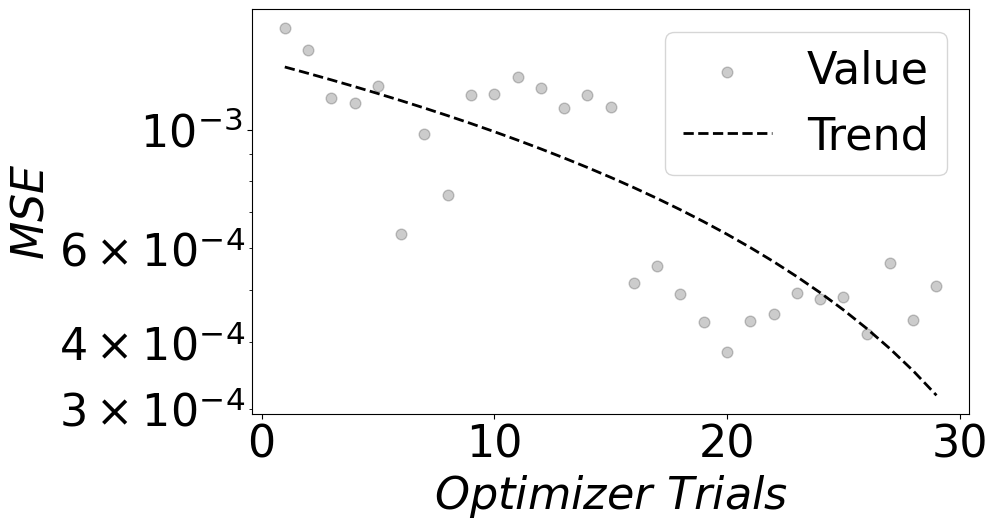}}

\caption{Hyperparameter optimization history of the AE--GCN model for each test case.}
\label{fig_trail_optimizer}
\end{figure}

Figure~\ref{fig_trail_optimizer} illustrates the optimization history of AE--GCN hyperparameters using Bayesian optimization. Each trial is represented by a set of transparent points indicating the \texttt{MSE} at the end of training. The dashed black line indicates the trend of error reduction during optimization. The graph underscores a continual decrease in error during the optimization, underscoring the efficacy of the tuning process in discovering the hyperparameters combination that minimizes \texttt{MSE} on the validation dataset.

A detailed computational cost analysis was conducted to evaluate the efficiency of the implemented AE--GCN model in comparison to the high--order approach, as outlined in Table~\ref{tab_cpu_time_comparison}. In CFD simulations, a single run typically consumes around 450 CPU hours, while generating the entire dataset demands roughly 31,500 CPU hours. Conversely, employing the ROM enables prediction for a single sample in approximately 1 second, resulting in a computational saving exceeding $99\%$. However, it is essential to consider the high computational cost associated with each high--fidelity simulation used for generating the dataset. Therefore, adopting a philosophy aimed at minimizing the amount of training data necessary for developing an accurate model is crucial.

The training process was executed on an Intel XEON W-2255 CPU with a NVIDIA RTX A4000 GPU, ensuring efficient utilization of computational resources.

\begin{table}[!b]
\small
\centering
\begin{tabular}{l c c c c c c}
\hline 
\hline 
\textbf{Test case} &  \multicolumn{2}{c}{\textbf{CFD (CPU hours)}} &  \multicolumn{3}{c}{\textbf{AE--GCN (GPU hours)}} \\
\cmidrule(lr){2-3}\cmidrule(lr){4-6}
  & \multicolumn{2}{c}{Simulation}   & Optimization & Training & Prediction \\
  & (70 runs) & (1 run)  & (30 trials) & (1 model) &  (1 sample) \\
\hline
BSCW & 35,000 &  500 &  27 &  1.4 &  0.0003 ($\sim{1s}$) \\
NASA CRM & 28,000 &  400 &  28 &  1.5 &  0.0003 ($\sim{1s}$) \\
\hline
\hline 
\end{tabular}
\caption{Computing cost comparison between AE--GCN model and CFD for the two test cases.}
\label{tab_cpu_time_comparison}
\end{table}

\bibliographystyle{elsarticle-num-names} 
\bibliography{cas-refs}

\end{document}